\documentclass[useAMS,usenatbib]{mn2e}

\usepackage{graphicx,alltt}
\usepackage{caption}

\usepackage{graphicx}
\usepackage{natbib} 
\usepackage{amsmath,amssymb}

\title[Composite nature of Dust-Obscured Galaxies (DOGs) at z$\sim$2-3 in the COSMOS field]{The composite nature of Dust-Obscured Galaxies (DOGs) at z$\sim$2-3 in the COSMOS field: I. A Far-Infrared View}
\author[Riguccini et al.]{L. Riguccini$^{1,2}$\thanks{E-mail:
riguccini@astro.ufrj.br}, E. Le Floc'h$^{3}$, J.R. Mullaney$^{4}$, K. Men\'endez-Delmestre$^{1}$, H. Aussel$^{3}$,   
\newauthor S. Berta$^{5}$, J. Calanog$^{6}$, P. Capak$^{7}$, A. Cooray$^{6}$, O. Ilbert$^{8}$, J. Kartaltepe$^{9}$, 
\newauthor A. Koekemoer$^{10}$, D. Lutz$^{5}$, B. Magnelli$^{5}$, H. McCracken$^{11}$, S. Oliver$^{12}$, I. Roseboom$^{13}$, 
\newauthor M. Salvato$^{14}$, D. Sanders$^{15}$, N. Scoville$^{16}$, Y. Taniguchi$^{17}$, E. Treister$^{18}$,  \\
$^{1}$Observat\'orio do Valongo, Universidade Federal do Rio de Janeiro, Ladeira do Pedro Ant\^onio 43, Sa\'ude, \\ Rio de Janeiro, RJ 20080-090, Brazil (riguccini@astro.ufrj.br)\\
$^{2}$CAPES/BJT Science Without Borders Postdoctoral Fellow, Brazil\\
$^{3}$Laboratoire AIM, CEA/DSM-CNRS-Universit\'e Paris Diderot, IRFU/Service d'Astrophysique, B\^at.709, CEA-Saclay, \\
91191 Gif-sur-Yvette Cedex, France \\
$^{4}$Department of Physics and Astronomy, Hicks Building, University of Sheffield, S3 7RH, U.K.\\
$^{5}$Max-Planck-Institut f\"ur extraterrestrische Physik, Postfach 1312, Giessenbachstrasse 1, 85741 Garching, Germany\\
$^{6}$Center for Cosmology, Department of Physics and Astronomy, University of California, Irvine, CA 92697, USA\\
$^{7}$Spitzer Science Center, 314-6 Caltech, Pasadena, CA 91125, USA\\
$^{8}$Laboratoire d'Astrophysique de Marseille, BP 8, Traverse du Siphon, 13376 Marseille Cedex 12, France\\
$^{9}$National Optical Astronomy Observatory, 950 N. Cherry Ave., Tucson, AZ, 85719, USA\\
$^{10}$Space Telescope Science Institute, 3700 San Martin Drive, Baltimore, MD 21218, USA\\
$^{11}$Institut d'Astrophysique de Paris, UMR7095 CNRS, Universit\'e Pierre et Marie Curie, 98 bis Boulevard Arago, 75014 Paris, France\\
$^{12}$Astronomy Centre, Department of Physics \& Astronomy, University of Sussex, Brighton BN1 9QH, UK\\
$^{13}$Institute for Astronomy, University of Edinburgh, Royal Observatory, Blackford Hill, Edinburgh EH9 3HJ, UK\\
$^{14}$Max-Planck-Institute f\"ur Plasma Physics, Boltzmann Strasse 2, Garching 85748, Germany\\
$^{15}$Institute for Astronomy, 2680 Woodlawn Dr., University of Hawaii, Honolulu, HI 96822, USA\\
$^{16}$California Institute of Technology, MC 105-24, 1200 East California Boulevard, Pasadena, CA 91125, USA\\
$^{17}$ Research Center for Space and Cosmic Evolution, Ehime University, Bunkyo-cho, Matsuyama 790-8577, Japan\\
$^{18}$Universidad de Concepci\'on, Departamento de Astronom\'ia, Casilla 160-C, Concepci\'on, Chile}
\begin{document}

\date{Accepted 2015 June 08.  Received 2015 May 15; in original form 2014 September 23.}

\pagerange{\pageref{firstpage}--\pageref{lastpage}} \pubyear{2002}

\maketitle

\label{firstpage}

\begin{abstract}
Dust-Obscured galaxies (DOGs) are bright 24$\mu$m-selected sources with extreme obscuration at optical wavelengths. 
They are typically characterized by a rising power-law continuum of hot dust (T$_D$ $\sim$ 200-1000\,K) in the near-IR indicating that their mid-IR luminosity is dominated by an an active galactic nucleus (AGN). 
DOGs with a fainter 24~$\mu$m flux display a stellar bump in the near-IR and their mid-IR luminosity appears to be mainly powered by dusty star formation. Alternatively, it may be that the mid-IR emission arising from AGN activity is dominant but the torus is sufficiently opaque to make the near-IR emission from the AGN negligible with respect to the emission from the host component. 
In an effort to characterize the astrophysical nature of the processes responsible for the IR emission in DOGs, this paper exploits Herschel data (PACS + SPIRE) on a sample of 95 DOGs  within the COSMOS field. We derive a wealth of far-IR properties (e.g., total IR luminosities; mid-to-far IR colors; dust temperatures and masses) based on SED fitting.
Of particular interest are the 24~$\mu$m-bright DOGs (F$_{24 \mu m} >$1mJy). They present bluer far-IR/mid-IR colors than the rest of the sample, unveiling the potential presence of an AGN. The AGN contribution to the total 8-1000$\mu$m flux increases as a function of the rest-frame  8$\mu$m-luminosity irrespective of the redshift. This confirms that faint DOGs (L$_{8 \mu m}<$ 10$^{12}$ L$_{\odot}$) are dominated by star-formation while brighter DOGs show a larger contribution from an AGN.
\end{abstract}


\section{Introduction}
\label{sec:intro}

The unprecedented sensitivity and angular resolution of the {\it
  Spitzer} Space Telescope at infrared (IR) wavelengths led to the
discovery of a new type of galaxy that is extremely faint in the optical ($\sim$\,22$<$R$<$27), yet bright at mid-infrared wavelengths
\citep{Houck:05,Dey:08,Fiore:08}. These sources, known to as ``Dust
Obscured Galaxies'' (hereafter, DOGs) in reference to the cause of
their faintness at optical wavelengths, have extremely red optical-to-IR colors ($f_{\nu}(24 \mu
m)/f_{\nu}(R)>982$). The incidence of DOGs is
relatively low: only 8\,$\%$ of 24~$\mu$m detected sources are DOGs, while $\sim$40\,\% of the sources in the 2\,deg$^{2}$ COSMOS field optical catalog have similar R-band magnitude [22.4-26.4]. However, their contribution to the total IR output of the Universe at $z\sim2$ where their source numbers peak
is estimated to be at least 30\% \citep{Riguccini:11}. 
This contribution increases to 50\,\% when considering the high-luminosity tail of their distribution at these redshifts \citep[i.e., $L_{\rm IR}>10^{12}$~L$_{\odot}$ ; e.g.,][] {Riguccini:11}. DOGs have {\it IR} luminosities
$>10^{11}$~L$_{\odot}$\, placing them in the LIRG and ULIRG\footnote{Luminous Infra-red Galaxies with 10$^{11}\,L_{\odot}<$L$_{IR} < 10^{12}$\,L$_{\odot}$ and Ultra-Luminous Infra-red Galaxies with  L$_{IR} > 10^{12}$\,L$_{\odot}$ \citep[e.g.,][]{Sanders:88a,Sanders:88b}} class of 
galaxies \citep[e.g.,][]{Dey:08,Bussmann:09,Riguccini:11}. Such luminosities require
significant amounts of dust-heating, most probably arising from
star-formation and/or high levels of nuclear
activity (i.e., active galactic nucleus or AGN). A number of recent
studies have split the DOG population along these lines: i.e., DOGs
showing a ``bump'' at 1.6~$\mu$m indicative of star-formation
\citep[][hereafter bump DOGs]{Farrah:08,Desai:09} and DOGs displaying a rising power-law
SED in the near- to mid-IR bands, suggesting a dominant AGN
\citep[][hereafter PL-DOGs]{Houck:05,Weedman:06}. Estimating the star-formation rate of the
latter has proved extremely difficult due to the dominant AGN
component washing out any host galaxy signatures.

The faintness of DOGs at optical wavelengths has made the characterization of their
physical properties particularly challenging. The launch of
{\it the Herschel Space Telescope} in 2009 provided a new window onto these galaxies that is
largely independent of dust obscuration, thereby giving us the
clearest view yet of these galaxies. The wavelengths probed by {\it
  Herschel} cover the peak of the spectral energy distribution
(hereafter, SED) of DOGs at the redshifts where their numbers are
highest (i.e., $1.5\lesssim z \lesssim3$). This allows us to accurately
constrain important properties, including the total IR luminosity as well as
dust temperature and mass. The aim of this work is to use the
combined diagnostic powers of both {\it Spitzer} and {\it Herschel}
observations to determine how these properties relate to the
dominant source of energy in these galaxies be it AGN, 
intense star formation or a combination of both. For this we use a sample of Spitzer/MIPS 24~$\mu$m-selected
DOGs (satisfying $F_{24 \mu m}> 0.08$~mJy) selected from the
COSMOS field \citep{Sco:07b}, and detected in all 5 {\it Herschel} bands. We calculate the
contribution from AGN and/or star-formation to the total energy output
of these galaxies via SED fitting and relate this to their dust
temperature and masses.  

The paper is organized as follows. Our data
are described in Sect.  \ref{sec:dat}, the far- to mid-IR colors of DOGs
sources are detailed in Sect. \ref{sec:colors}. The SED-fitting
procedure used and the results obtained are described in
Sect. \ref{sec:sed_fitting}. In Sect. \ref{sec:dust} we present the 
model and results on the dust temperature and mass of our
DOG sample and discuss if the presence of AGN signatures induce a
particular trend in the T$_{dust}$ distribution. We discuss
our results and present our conclusions in
Sect. \ref{sec:ccl}. Throughout this paper we assume a $\Lambda$CDM
cosmology with H$_{0}$=70 km s$^{-1}$, $\Omega_m$ = 0.3, and
$\Omega_{\Lambda}$ = 0.7. Unless otherwise specified, magnitudes are given in the AB system.

\section{Data}
\label{sec:dat}

The sample of DOG sources is selected from the deep {\it
  Spitzer}/MIPS observations of the 2 deg$^{2}$ COSMOS field
\citep{Sanders:07}. Our starting point are the 24~$\mu$m\ detected
sources from the catalogue described in \citet{Emeric:09} \citep[see also][]{Riguccini:11}. We note that other studies further require
  a source to satisfy $f_{24 \mu m}>300~{\rm mJy}$ in order to classify it as a DOG \citep[e.g.,][]{Dey:08}. In
  this study we consider all sources satisfying $f_{\nu}(24 \mu
  m)/f_{\nu}(R)>982$ as DOGs. Furthermore, DOG studies focussing on heavily-obscured AGNs \citep[e.g.,][]{Fiore:08,Fiore:09,Treister:09} also impose an additional R-K$>$4.5 (vega) cut.

\subsection{COSMOS observations}
\label{sec:cosmos}

COSMOS is a wide-area equatorial field with deep coverage at all
wavelengths spanning radio to X-rays
\citep{Hasinger:07,Schinnerer:07,Elvis:09}. Crucial for this study is
the deep IR coverage of this field, particularly at mid- to far-IR wavelengths by the {\it Spitzer Space Telescope} with the MIPS instrument \citep{Emeric:09}
and, more recently, with PACS \citep{Poglitsch:10} and SPIRE \citep{Griffin:10} onboard {\it
  Hershel} \citep{Pilbratt:10}. 

The extensive UV to
near-IR coverage of COSMOS
\citep[e.g.,][]{Taniguchi:07,Capak:07} allows for precise photometric
redshifts (hereafter, photo-z) to be derived for extragalactic sources within this field. For the photo-zs used in this work, we use an updated\footnote{version 1.8: the main improvements compared to \citet{Ilbert:09} reside in relying on the median of the PDF to define the ``best" photo-z, instead of the minimum $\chi^2$} version of the photometric redshift catalog
of \citet{Ilbert:09} that provides photo-zs for 1\,400\,237 i$^+$-detected
sources among the 2\,017\,800 sources of the COSMOS photometric
catalog. These redshifts are obtained with an unprecedented accuracy,
with a dispersion of $\sigma_{\Delta z /
  (1+z)}=0.012$ for sources satisfying $i^{+}_{AB}<$24 and z$<$1.25.
More relevant to this study $-$ where we focus on dusty 24$\mu m$-selected
sources that are very faint at optical wavelengths $-$ is their
comparison with the optically-faint spectroscopic sample from the
z-COSMOS survey \citep{Lilly:07} where \cite{Ilbert:09} report a
dispersion of only $\sigma_{\Delta z /
  (1+z)}$\,=\,0.06 for sources with 23$<i^{+}_{AB}<$25 at 1.5$<z<$3.
Given their accuracy for faint sources, we use these photo-zs for our
24~$\mu$m\ sources, matching their optical counterparts following the
procedure outlined in \citet{Emeric:09} and \citet{Riguccini:11}. We briefly describe this approach in the following subsection.

\subsection{The far-IR counterparts of 24~$\mu$m-selected sources}

Our 24~$\mu$m parent sample (from \citet{Emeric:09} and \citet{Riguccini:11}) contains
29\,395 sources detected at 24~$\mu$m with F$_{24 \mu m}>~$80~$\mu$Jy
over a total area of 1.68 deg$^{2}$, which excludes regions
contaminated by bright, saturated objects. In the interest of focussing on the sources' star-formation
histories, we exclude X-ray detected AGNs down to a flux limit of S$_{0.5-2\,kev}$\,=\,5\,$\times$\,10$^{-16}$\,erg\,cm$^{-2}$\,s$^{-1}$  based on AGN catalogs from \citet{Brusa:07,Brusa:10} and \citet{Salvato:09}.

We limit our counterpart identification to 24~$\mu$m sources with a 3-$\sigma$ {\it PACS} detection at 100~$\mu$m and 160~$\mu$m. SPIRE fluxes will be used for a subset of our sample.
The COSMOS field was observed as part of the
 PACS Evolutionary Probe \citep[PEP,][]{Lutz:11} and the Herschel Multi-tiered Extragalactic Survey \citep[HerMES,][]{Oliver:12} campaigns (i.e., PACS 100 \& 160 $\mu$m and SPIRE 250,
350 \& 500 $\mu$m respectively). The catalogs provided by PEP and HerMES calculate source fluxes in each of these 5
bands by performing PSF fitting at the positions of
the 24~$\mu$m-detected sources from \citet{Emeric:09}. One of the key benefits of using such 24~$\mu$m\
``priors'' as opposed to generating blind catalogues, is that it helps
with deblending, which is particularly problematic at the longer {\it Herschel} wavelengths. 
The HerMES catalog was built following the method presented in
\citet{Roseboom:10}, based on the 24~$\mu$m position priors from
\citet{Emeric:09}. The PEP catalog was obtained using the same {\it 24~$\mu$m priors} \citep{Berta:11}. The reliability and the completeness of the PACS
and SPIRE COSMOS catalogs are detailed in \citet{Lutz:11} and
\citet{Oliver:12}, respectively. We identify a total of 6\,029 24~$\mu$m-detected sources with a $>$3-$\sigma$ detection in the PACS-bands with F$_{100 \mu m}>~$3.1~mJy and F$_{160 \mu m}>~$6.3~mJy (see Table\,\ref{tab:nb_sources}).
We match these to the
catalogue of optical sources from \citet{Ilbert:09} in order
to obtain their photometric redshifts. 

Given the much higher density of sources detected at optical wavelengths in COSMOS \citep{Capak:07} compared to those detected with MIPS, a direct cross-correlation between the 24$\mu$m-selected catalog and the optical observations could lead to a large number of spurious associations with optically-detected galaxies randomly-aligned close to the line of sight of the MIPS sources. To minimize this, we first matched our 24$\mu$m catalogue to the K-band catalogue
of \citet{McCra:10}, employing a matching radius of 2" and following
the same procedure described in \citet{Emeric:09} and
\citet{Riguccini:11}. Of the 6\,029 sources in our sample, 5858 were
found to have a K-band counterpart. In an attempt to reduce the number
of non-matches, we also matched our 24$\mu$m catalogue to the IRAC-3.6~$\mu$m catalog from \citet{Capak:07},
adopting the same 2'' matching radius. This led to 34 additional
matches, increasing to 5892 the number of MIPS-24~$\mu$m$+$Herschel
sources with either a K-band or an IRAC-3.6~$\mu$m counterpart. These
5892 sources were then matched to the updated version of the
i$^{+}$-band selected catalog of photometric redshifts from
\citet{Ilbert:09} using a matching radius of 1". Of the 5892 sources,
5768 had i$^{+}$-band counterparts and associated photometric
redshifts, leaving 261 sources among the 6\,029 24~$\mu$m$+$Herschel
sources (i.e., $\approx4\%$) without photometric redshifts. These sources were
excluded from any further analyses.

\subsection{The PACS-DOGs sample}
\label{sec:sample}

\begin{figure} 
 \resizebox{1.\hsize}{!}{\includegraphics{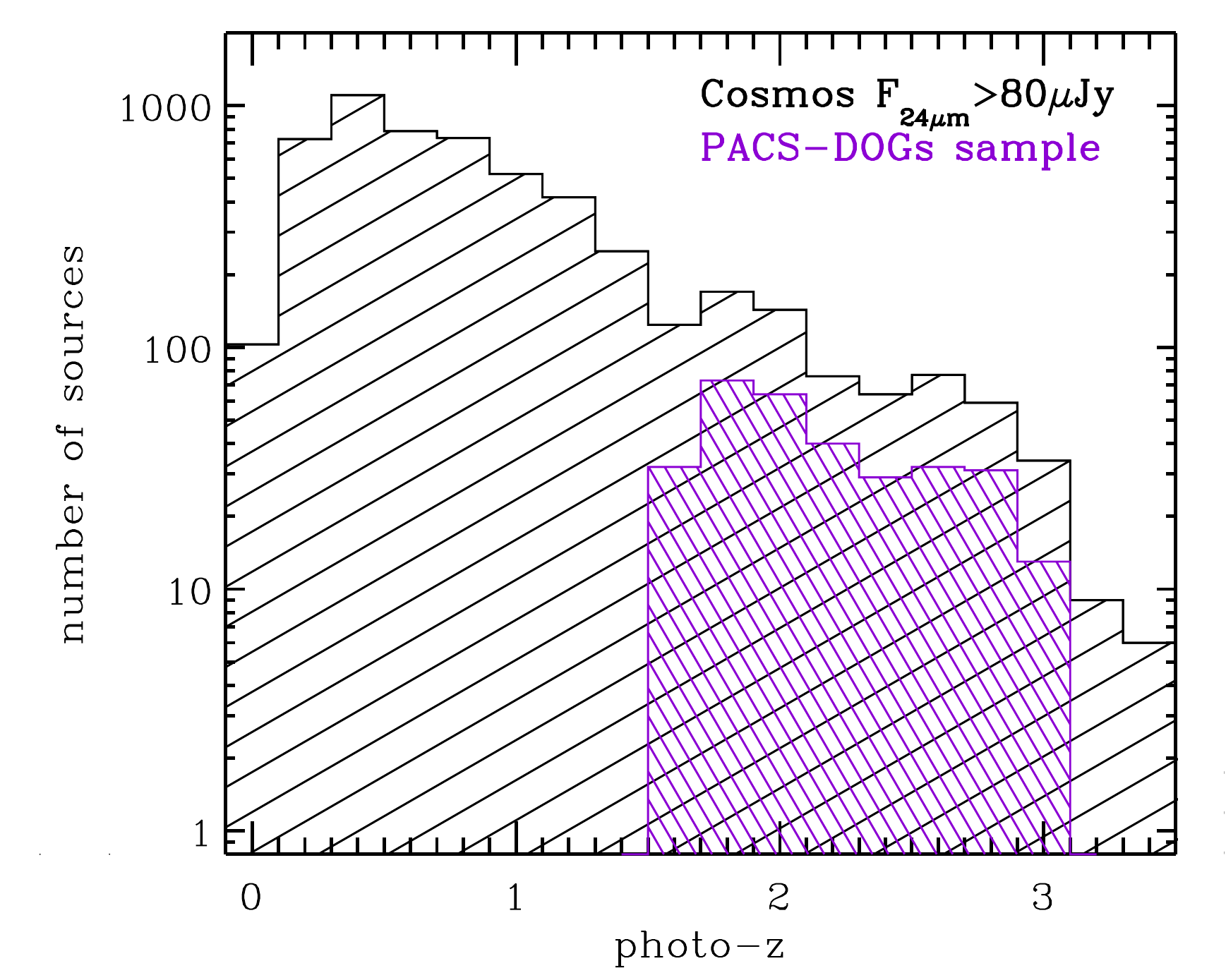}}
 \caption{Photometric redshift distribution of the 24$\mu$m sources from \citet{Emeric:09} in black and of the PACS-DOGs from this  work in purple.}
 \label{fig:dist_z_pacs_sources}
\end{figure}

Our DOGs sample was selected from the 5\,892 sources selected at 24~$\mu$m with a 3-$\sigma$ detection in at least one of the PACS bands (100 or 160~$\mu$m). The DOGs criterion introduced by \citet{Dey:08} is based on the following: $F_{24 \mu m}/F_{R}>$982 and $F_{24 \mu m}>$300
$\mu$Jy, where the latter is a direct consequence of the depth of the MIPS imaging in the Bootes field. Considering that the source-extraction performed by \citet{Emeric:09} reaches a completeness of $\sim$90\% with $F_{24 \mu m}>$80 $\mu$Jy, we extend the DOGs 24 $\mu$m-flux cut down to 80 $\mu$Jy.
The R-band magnitudes used in this work are from \citet{Ilbert:09} based on observations with the Subaru telescope by \citet{Capak:07}; these include a correction for Galactic extinction -- not applied in \citet{Capak:07} --  and reach a limiting magnitude of M$_{R}\,>$\,17.5\footnote{We note that 24 $\mu$m-selected sources with no R-band detection are also considered DOG sources.}. 

The sample contains 57 and 138 DOGs detected in only one of the PACS bands at 100~$\mu$m and 160~$\mu$m, respectively, while 119 DOGs are detected in both PACS bands. This amounts to a total of
314 PACS-detected DOG sources (cf Table\,\ref{tab:nb_sources}) with M$_{R}\,>$\,23.4, F$_{100 \mu m}>$3.4\,mJy and F$_{160 \mu m}>$7.8\,mJy within the redshift range 1.5$<z<$3 (see Fig.\,\ref{fig:dist_z_pacs_sources}), where the DOGs criterion is the most efficient \citep[e.g.,][]{Dey:08,Bussmann:09,Bussmann:12,Riguccini:11}. We will use this sample in Sec.\,\ref{sec:colors}.

\subsection{The {\it Herschel}-DOGs sample}
\label{sec:sample2}

\begin{figure} 
 \resizebox{1.\hsize}{!}{\includegraphics{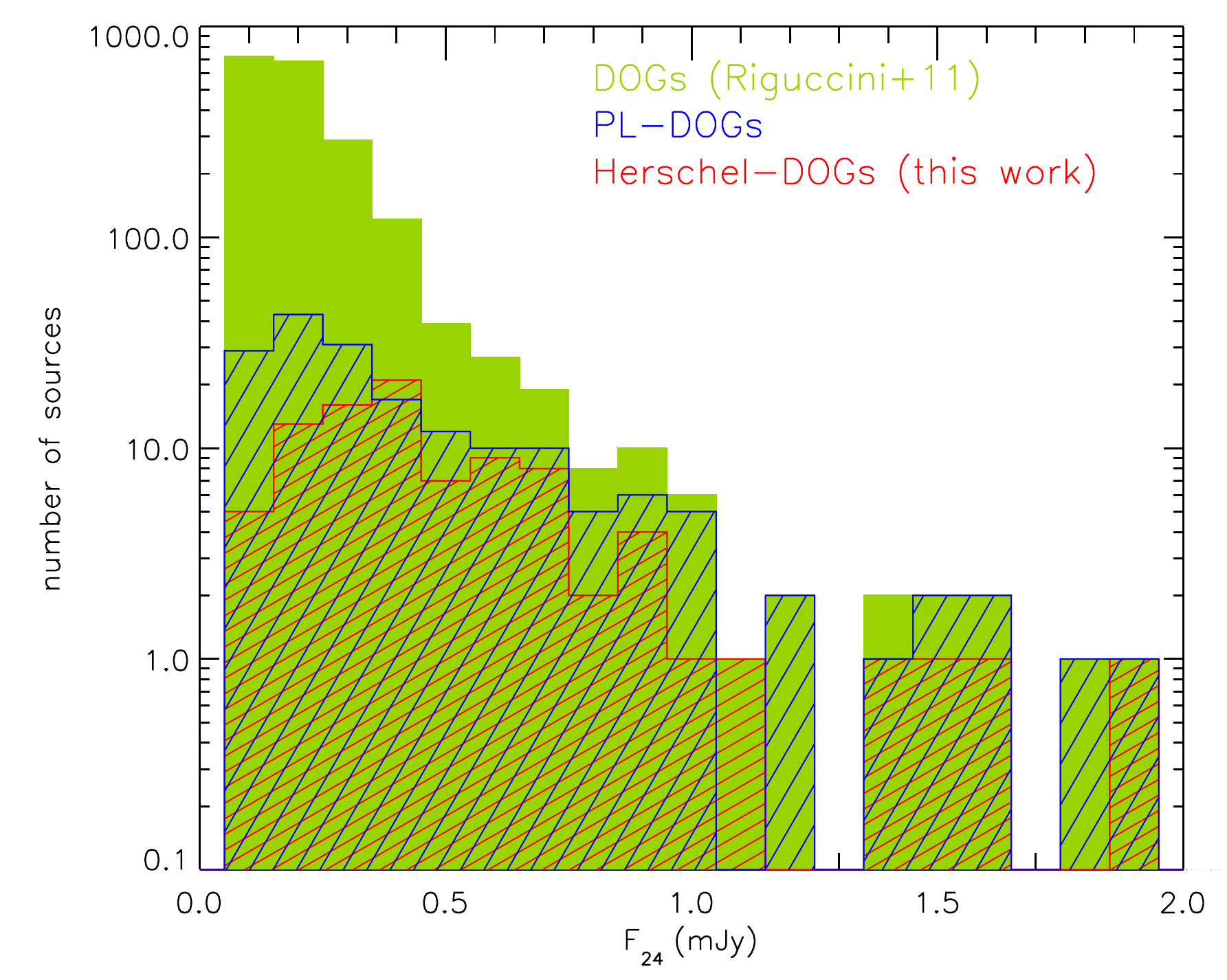}}
 \caption{Distribution of 24~$\mu$m flux for the DOG parent sample ($\sim$2100 sources) from \citet{Riguccini:11} (green), PL-DOGs (blue) and the {\it Herschel}-DOGs from this work (red). The {\it Herschel}-DOGs distribution peaks at higher 24~$\mu$m fluxes ($\sim$0.36\,mJy according to a Gaussian fit), compared to that of the whole DOGs sample ($\sim$0.14\,mJy). The distribution of the PL-DOGs selected from \citet{Riguccini:11} also peaks slightly higher ($\sim$0.22\,mJy) than the whole DOGs population distribution and is also more inclined to select 24~$\mu$m bright sources. See Sect. \ref{sec:sample2} for details.}
 \label{fig:ks_test}
\end{figure}

\begin{figure} 
 \resizebox{1.\hsize}{!}{\includegraphics{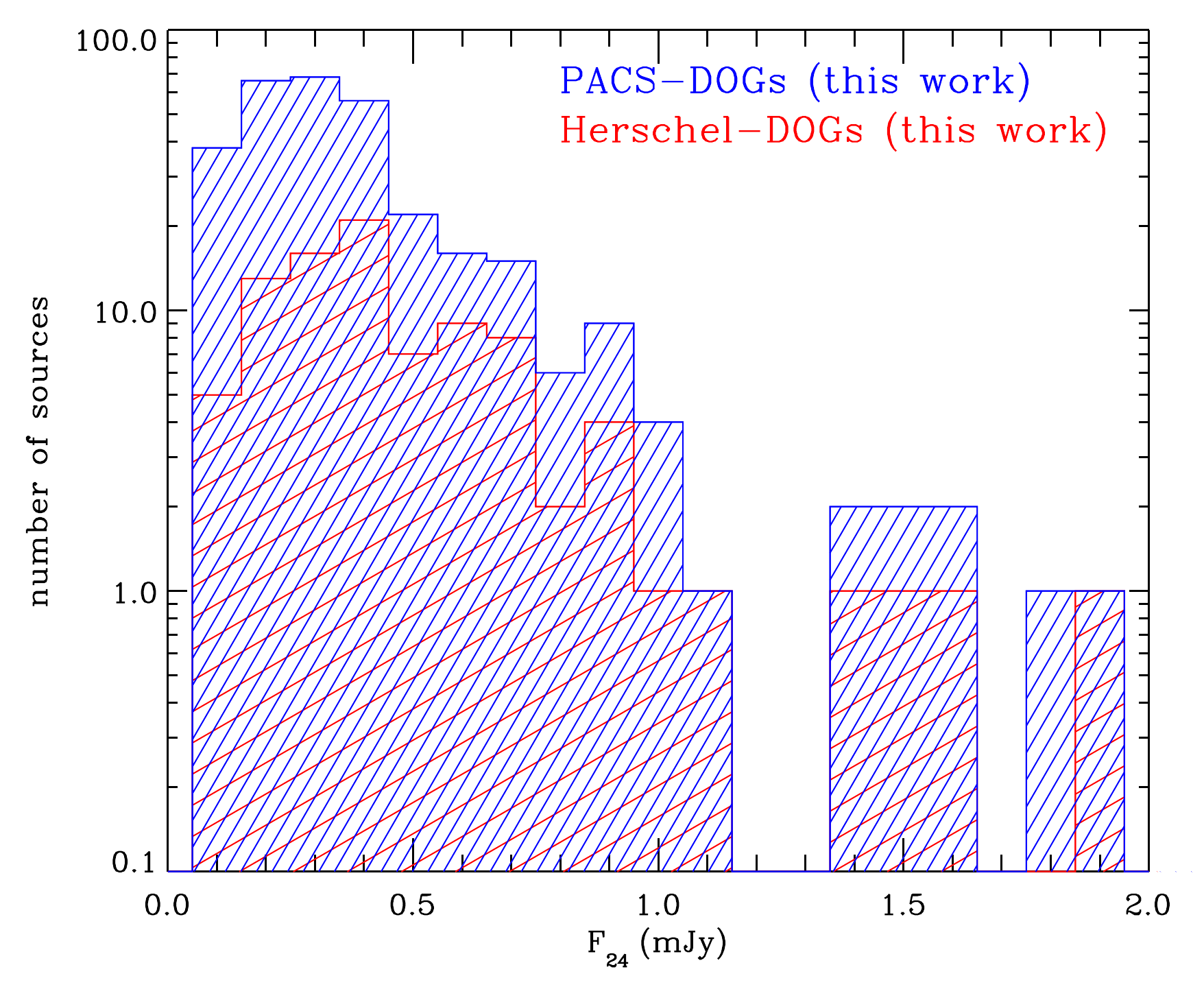}}
 \caption{Distribution of 24~$\mu$m flux of DOGs detected in at least one PACS-band (314 sources) in blue and the same {\it Herschel}-DOGs distribution than in Fig.\,\ref{fig:ks_test} in red.}
 \label{fig:ks_test_PACS}
\end{figure}

Considering that we seek to undertake SED-fitting across the mid- to far-IR wavelength range, we further define a sub-sample of PACS-DOGs detected in all 5 Herschel bands ; this allows for a better constraint on the peak of the DOGs'
SEDs.
To achieve this we matched the PACS-DOGs sample with the SPIRE catalog \citep{Roseboom:10}. This results in 95 Herschel-detected DOGs\footnote{All 95 Herschel-detected have R-band detections}, i.e. detected in the 5 {\it Herschel} bands (see Table\,\ref{tab:nb_sources}). 

The fluxes in the PACS bands are obtained for all sources with a $>3$-$\sigma$ detection. Although the SPIRE catalog reaches a 3-$\sigma$ limit of $\sim$\,10
mJy, $\sim$ 12 mJy and $\sim$ 15 mJy at 250, 350 and 500\,$\mu$m
respectively, the corresponding 3-$\sigma$ extragalactic confusion limits are 14.4 mJy, 16.5 mJy and 18.3 mJy \citep{Nguyen:10}. In the SED-fitting procedure we are cautious \citep{Magnelli:12a}  when including fluxes that are lower than the 3-$\sigma$ extragalactic confusion limits and use them merely as upper limits.

Among our sample of 95 Herschel-detected DOGs, 40 have their fluxes above the 3-$\sigma$
threshold only for the 250 and 350 $\mu$m bands, 20 merely for the 250$\mu$m band, one DOG for the 350 and 500 $\mu$m bands and another DOG solely for the 350 $\mu$m; 9 of the Herschel-detected DOGs have all SPIRE fluxes below the 3-$\sigma$
threshold. We quote upper limits for all of these cases. Only 24
sources have fluxes above the 3-$\sigma$ limit in the three SPIRE
bands. 

We acknowledge that imposing a detection in the 5 {\it Herschel} bands will impart a bias towards the brightest and reddest IR sources in
our sample. This is shown in Fig.\,\ref{fig:ks_test}, where the distribution of {\it Herschel}-DOGs peaks at higher 24~$\mu$m fluxes  than that of the DOGs parent sample from \citet{Riguccini:11}.
Of particular interest is to note that although the faintest DOGs (F$_{24\mu m}<$0.4\,mJy) are missed by the {\it Herschel}-selection, beyond F$_{24\mu m}>$0.4\,mJy the Herschel-DOGs distribution is very similar to that of the DOG parent population (see Fig.\,\ref{fig:ks_test}). On the other hand, PL-DOGs -- known to be mainly AGN dominated \citep[e.g.,][]{Bussmann:09} -- present a significantly stronger bias towards $24\mu$m-bright sources: not only they have a $24\mu$m flux distribution that peaks slightly higher than the whole DOGs population distribution, but their selection represents  60\%  of the DOG population with F$_{24\mu m}>$1\,mJy, compared to merely a 10\% at F$_{24\mu m}=$0.3\,mJy \citet{Bussmann:09}. 

In this paper, two samples of DOGs are used. To study the IR colors of DOGs (Section\,\ref{sec:colors}) we use only PACS data and thus base our analysis on the 314 PACS-DOGs, in an effort to improve our statistics. For the remainder of our study we restrict our analysis to the 95 Herschel-detected DOGs, noting that both samples probe the same DOG population, as illustrated by their 24~$\mu$m flux distributions on Fig.\,\ref{fig:ks_test_PACS}. We focus our study on the differences observed between mid-IR bright DOGs (F$_{24\mu m} >$ 1\,mJy) and DOGs with more moderate fluxes ($\sim$ 0.2\,mJy$<$F$_{24\mu m} <$ 1\,mJy).

\subsection{The reliability of the photometric redshifts for the DOGs sample}
\label{sec:photo-z}

The high accuracy of the photometric redshifts for sources of the COSMOS
catalogue (see Sect.\,\ref{sec:cosmos}) make them highly reliable for
statistical studies on large (i.e., $>$2\,000) samples
\citep[e.g.,][]{Riguccini:11}. However, because in this study we focus
on significantly smaller numbers, we require particularly robust
redshift measurements for each source. To ensure this, we checked the
distribution of the probability density function (PDF) of the
photometric redshift for each source and divided our sample 
into three categories, according to the photometric-redshift reliability. The categories are the following: (1) sources have a single, secure
photo-z, for which the PDF has a gaussian shape with a single peak (39 sources); (2) those
with multiple potential photo-zs, for which either the PDF's peak is spread over a wider range of redshifts ($\Delta z \sim$ 0.2)
  or it
includes a lesser peak which may correspond to another photometric
redshift (24 sources); (3) those flagged as presenting inaccurate
photo-zs, because their PDF shows clear multiple peaks of similar
strength (32 sources).  For sources in the last two categories, the
photo-z is set to the highest peak value of the z-distribution
and in the case of multiple peaks we keep the value of the subsequent
peaks as secondary options. To get the most reliable and accurate fit to the SEDs of our sources, we use all of these potential photo-z values.
We make a special note that six DOGs in our sample have a confirmed COSMOS spectroscopic redshift as part of the Fiber Multi Object Spectrograph (FMOS) spectroscopic redshift catalog (Kartaltepe et al., in prep.) which we use in our analysis for higher accuracy.

\begin{table}
\caption{Number of sources} 
\centering
    \begin{tabular}{ | l | p{7cm} |}
    \hline
     6\,029 & 24~$\mu$m-sources with a 3-$\sigma$ PACS detection (at 100~$\mu$m and 160~$\mu$m) \\ \hline
 5\,892 & sources from the previous sample with a photo-z \\ \hline
 314 & DOGs (i.e., $F_{24 \mu m}/F_{R}>$982) with $F_{24 \mu m}>$80$\mu$Jy and 1.5$<z<$3 and with a 3-$\sigma$ detection in one or the two PACS-bands \\
 &  Sample used for the far-IR/mid-IR color analysis [Sect.\,\ref{sec:colors}]
  \\ \hline
 95 & DOGs with $F_{24 \mu m}>$80$\mu$Jy and 1.5$<z<$3 and with a 3-$\sigma$ detection in the 2 PACS bands and with a detection (potentially $>$ 3\,$\sigma$) in the 3 SPIRE bands \\
 &  Sample used for the remainder of the paper
\\ \hline
     \end{tabular}
\label{tab:nb_sources}
\end{table}

\section{Results: far-IR/mid-IR colors of extremely mid-IR bright DOG sources}
\label{sec:colors}

The DOG sources are not only an extreme sub-sample of ULIRGs but also represent a mix between sources
dominated by star-formation and those dominated by AGN activity
\citep[e.g.,][]{Houck:05,Fiore:08,Fiore:09,Bussmann:09,Melbourne:12}. 
 In this paper we seek to quantify the AGN contribution of these sources and study the evolution of this contribution with respect to other galaxy properties, including redshift, the 8\,$\mu$m rest-frame luminosity, total IR luminosity, dust temperature and dust mass.

Studies in the past years have explored the PL- and bump-DOGs population \citep[e.g.,][]{Pope:08,Melbourne:09}. It has been well established that PL-DOGs have an AGN contribution to their near-IR emission and that their far-IR emission is most likely dominated by star formation\citep{Calanog:13}. 
In this paper we aim to gauge the AGN contribution of these DOGs using {\it Herschel} far-IR data.
 
\begin{figure*} 
 \resizebox{0.49\hsize}{!}{\includegraphics{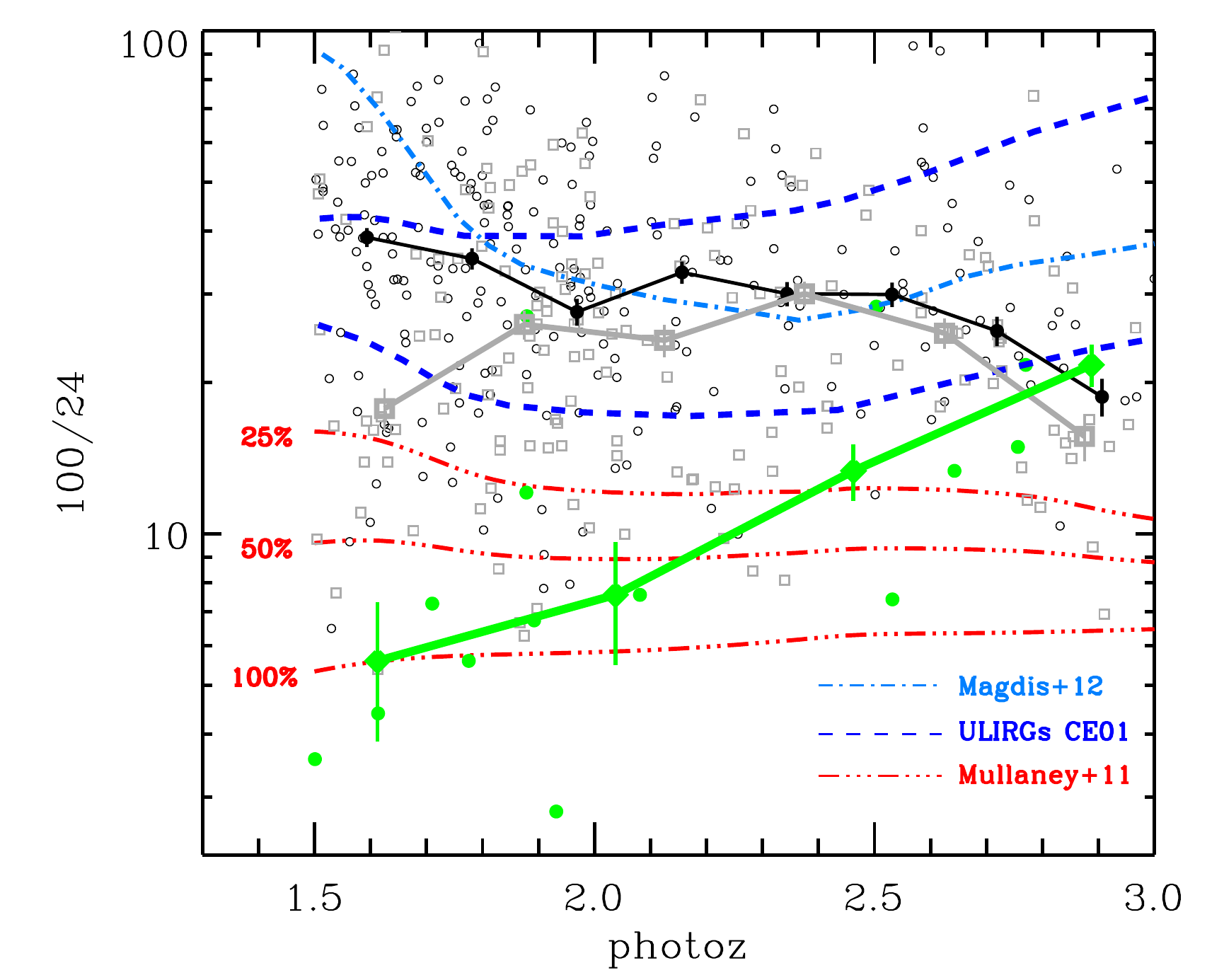}}
 \resizebox{0.49\hsize}{!}{\includegraphics{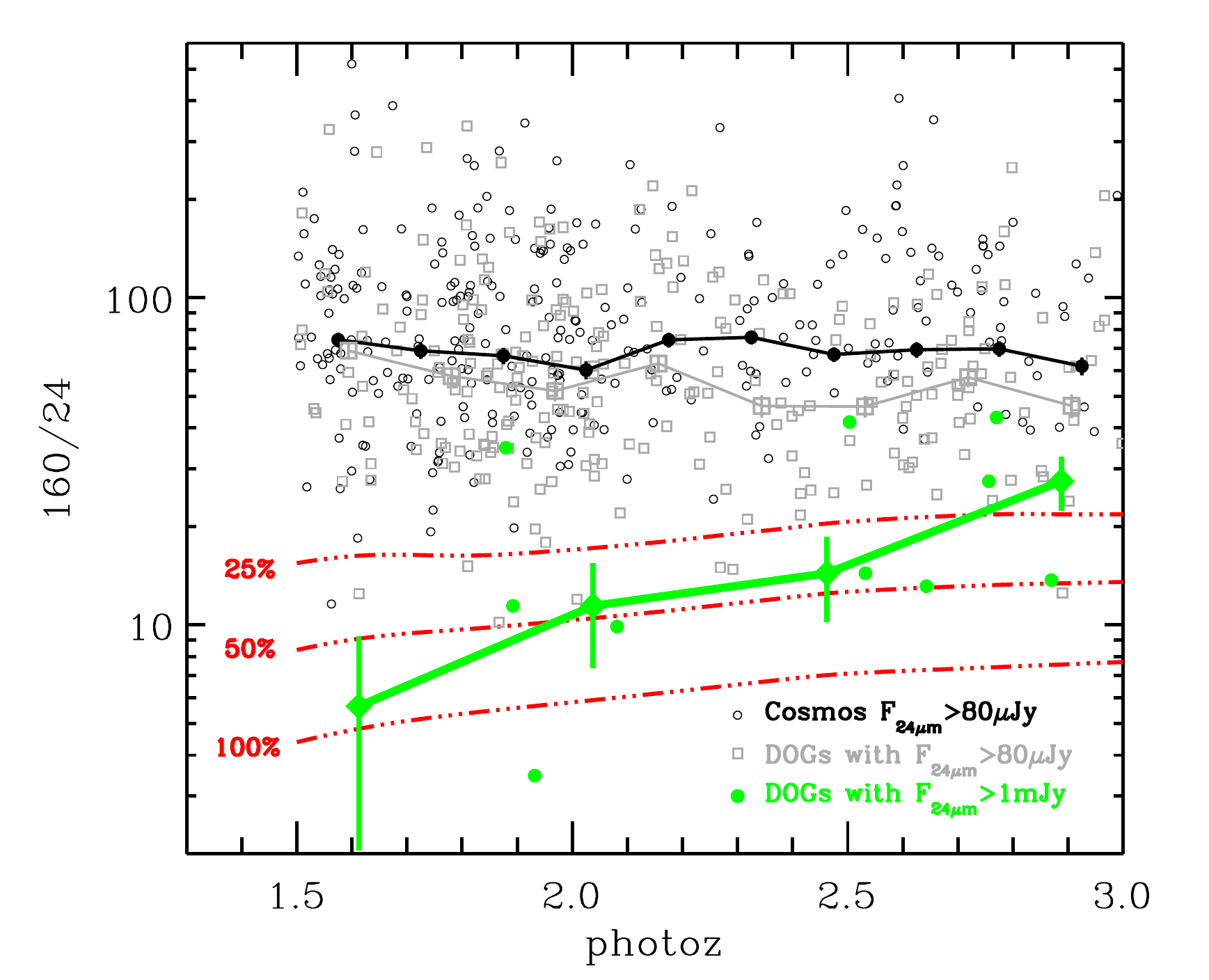}}
 \caption{{\it Left:} Distribution of 100$\mu$m/24$\mu$m color as a function of redshift for all the 24 $\mu$m-selected sources detected at 100 $\mu$m (black open circles), the 176 DOGs detected at 100 $\mu$m with $F_{24 \mu m}>$80 $\mu$mJy (grey open squares) and with $F_{24 \mu m}>$1 mJy (green filled circles). The black, grey and green solid curves, from top to bottom, represent the median of all the 24 $\mu$m-selected sources with F$_{24 \mu m}> $80 $\mu$Jy, the median of the 24~$\mu$m-sources selected as DOGs with $F_{24 \mu m}>$80 $\mu$mJy and the median of the brightest DOGs sources (i.e., with F$_{24 \mu m}> $1 mJy), respectively. 
Errors on the median are calculated as quadratic propagation of uncertainty.
 We show the expected flux ratios (red dashed-triple dotted tracks) in the case that 25, 50 and 100\% of the flux in the 100 $\mu$m band result from an AGN component \citep{Mullaney:11,Mullaney:12}. We also include for comparison the star-forming ULIRG CE01 templates for an IR luminosity of 10$^{12}$ L$_{\odot}$ (bottom blue-dashed line) and 10$^{12.5}$  L$_{\odot}$ (top blue-dashed line), as well as the template from \citet{Magdis:12} (light blue dotted-dashed line).
The observed PACS/24 colors of the bulk of the 24~$\mu$m sources and that of the DOG sources are consistent with these ULIRG templates. {\it Right:} Distribution of $160\mu m/24\mu m$ color as a function of the redshift for all the 24 $\mu$m-selected sources detected at 160 $\mu$m (black open circles) and the 257 DOGs detected at 160~$\mu$m. The colors and tracks are the same as on the left panel. For clarity, we do not overplot the CE01 templates on the right panel as they would give the same results than for the 100/24 color.}
 \label{fig:color_PACS_MIPS}
\end{figure*}

 As an initial, crude assessment of the dominant process responsible for producing most of
the IR output in DOGs (i.e., AGN vs. star formation), we first
consider the far- to mid-IR colors (hereafter FIR/MIR) of our sample \citep[e.g.,][]{Mullaney:12}. Fig.\,\ref{fig:color_PACS_MIPS} shows the 100~$\mu$m/24~$\mu$m
and 160~$\mu$m/24~$\mu$m color distributions for our
sample of PACS-detected DOGs as a function of redshift; we include
all 24~$\mu$m\ -detected COSMOS sources for comparison.  We see no noticeable trend  for the DOGs
sample at z$\ge$2; the curves shown for the 100/24 and 160/24 median color evolution with redshift seem to follow the same evolution than that of
the whole 24~$\mu$m-detected sample. However, at lower redshifts the DOGs
display a steeper evolution, with bluer 100/24 colors than the non-DOG 24$\mu$m-detected sources.

We find that the FIR/MIR distribution of both the bulk of the 24~$\mu$m comparison sample as well as the
majority of our DOGs are well represented by the star forming templates from \citep[][hereafter CE01]{Chary:01} with IR luminosities L$_{IR}$ = 10$^{12-12.5}$ L$_{\odot}$. This is to be expected given that IR-selected galaxies at z$>$1 tend to be of the LIRG or ULIRG class \citep[e.g,][]{Emeric:05,Magnelli:09}.  
 We emphasize that although recent work has shown that CE01 local ULIRG SEDs are not good fits to z$\sim$2 star-forming galaxies with similar IR luminosities \citep[e.g.,][]{Elbaz:10,Elbaz:11,Nordon:10,Nordon:12}, the CE01 ULIRGs templates are good fits to our DOGs; 17 of our sources are fit using these templates with a $\chi^2 <$5 (Sec.\,\ref{sec:sed_fitting}).

We consider the particular case of the brightest DOGs in our sample (i.e.,
$F_{24 \mu m}>$1 mJy) and find that they show a particular behavior in their FIR/MIR colors as a function of redshift:
the brightest DOGs in our sample show significantly bluer
PACS/24~$\mu$m colors than the general 24\,$\mu$m-detected population (i.e, with F$_{24\mu m}>$80 $\mu$Jy). We discard the possibility  that a variation in the Photodissociation regions (PDR) component and/or variation in the intensity of the field is responsible for the bluer color of these bright DOGs, by comparing to the 100~$\mu$m/24~$\mu$m colors derived from the templates of \citet{Magdis:12} (see Fig.\,\ref{fig:color_PACS_MIPS}). These SED templates are based on stacked ensembles at different redshift intervals, considering the varying radiation field and PDR contribution to ULIRGs as a function of redshift; for our study we rely on their starburst-dominated templates at the two relevant redshift intervals: $1.75<z<2.25$ and $2.27<z<3.0$.
We compare to the FIR/MIR colors of AGN/galaxy composites -- using the intrinsic AGN SED of
\citet{Mullaney:11} and assuming different AGN contributions (25, 50, 100\%) at
100~$\mu$m and 160~$\mu$m -- and find significant similarities, suggesting that the brightest DOGs have a
significant AGN contribution. As we consider higher redshifts, the median FIR/MIR color of these bright DOGs point towards a
lower fraction of the AGN contribution, 
consistent with the SFRs of galaxies of similar mass increasing
with redshift \citep[e.g.,][]{Brinchmann:04,Daddi:07,Pannella:09,Magdis:10}. We check the validity of these trends in the
following section by looking at the AGN contribution (based on SED-fitting) as a function of the 24~$\mu$m flux.

\section{AGN contribution to the total rest-frame 8-1000$\mu$m luminosity in DOGs}
\label{sec:sed_fitting}

Based on FIR/MIR colors, bright DOGs likely contain an AGN component, contributing partly or even dominating their IR luminosity. To have a better understanding of these sources and to have a global view of their stage in the evolution of the galaxies, it becomes important to know the exact contribution of a potential AGN to their total 8-1000$\mu m$ rest frame luminosity. In this section we present our method to determine the potential contribution of an AGN component to these DOG sources and show our results on the variation in AGN contribution with the 24~$\mu$m flux.

\subsection{Method: SED-fitting procedure}
\label{sec:proc}

Studying the SED of a galaxy provides insights to the physical nature of the underlying continuum source and can unveil the presence of an AGN. The impact that an AGN contribution has on the shape of the SED is distinct from that of dust heated by star-forming activity.
However, deriving the SED of a galaxy is not an easy exercise especially in the case of an AGN where the imprints of the host galaxy is always present. In our study we use the IDL-based SED-fitting procedure {\it DecompIR}, detailed in \citet{Mullaney:11}. Combining a set of five starburst templates and an average AGN template, this approach is aimed at fitting the IR photometry of composite galaxies and to measure the AGN contribution to their total IR output.
 A $\chi^{2}$ method is used to know which combination of these templates best fits the data; i.e., the combination with the lowest associated $\chi^{2}$ value is adopted as the best fit. The validity of this procedure as an accurate way to determine the AGN contribution to the total IR output of composite galaxies has been verified by several tests lead by \citet{Mullaney:11}, including a comparison with alternative measures of the AGN contribution (e.g., emission line diagnostics). Although there are significant uncertainties associated to the precise AGN contribution to an individual galaxy, this approach is adequate from a statistical point of view (i.e., large samples, average SEDs). 
 
 We apply the {\it DecompIR} procedure to our sample of DOGs sources. Nonetheless, considering that DOGs have ULIRG-class luminosities \citep[e.g.,][]{Bussmann:09,Riguccini:11}, we add two ULIRGs templates with $L_{IR}=10^{12} L_{\odot}$ and $L_{IR}=10^{12.5} L_{\odot}$ to the set of starburst templates from \citet{Mullaney:11} to fully cover the luminosity range of our sample. The ULIRG templates are taken from CE01 as they build their library from SEDs and models that only take into account star formation activity. The median PACS/MIPS colors of the bulk of the DOGs sample are well represented by these ULIRGs templates (see Fig.\,\ref{fig:color_PACS_MIPS}), motivating their use as part of our SED-fitting procedure. To determine the AGN contribution to the IR luminosity of our sources, we first derive a best SED-fit with templates based only on star-forming sources. If the star-forming templates do not provide a satisfactory result, an AGN component is added and the SED-fitting proceeds with a composite spectra. We consider the AGN component as a reasonable option only if it improves the $\chi^2$ of the fit by at least 50\%.

We implement our SED-fitting procedure to each DOG source in our sample. Each of the 95 sources are first fit by a star-forming component only and then by a composite spectra when the $\chi^2$ from the star-forming fit is $>$\,20. 
In the case of sources with a less-accurate photo-z, the source is fit with a star-forming template for all possible photo-zs obtained from the PDF (see Sect.\,\ref{sec:photo-z}). If none of these fits are suitable, we add an AGN component and the fits for each possible photo-z are performed once again.
 Our method is robust in fitting most of our sources (90\%). Out of 95 DOGs, 71 are fit with a star-forming galaxy template only, 15 require an AGN component to the fit, and in 9 cases, no reliable fit was obtained either using a star-forming-only template nor a composite spectra.
The failure of successfully fitting these sources could be due to the fact that we cannot reproduce the SED of these sources; this is most likely due to a wrong redshift, even after probing the different possibilities indicated by the PDF.

\begin{figure*} 
 \centering
 \resizebox{1.\hsize}{!}{\includegraphics{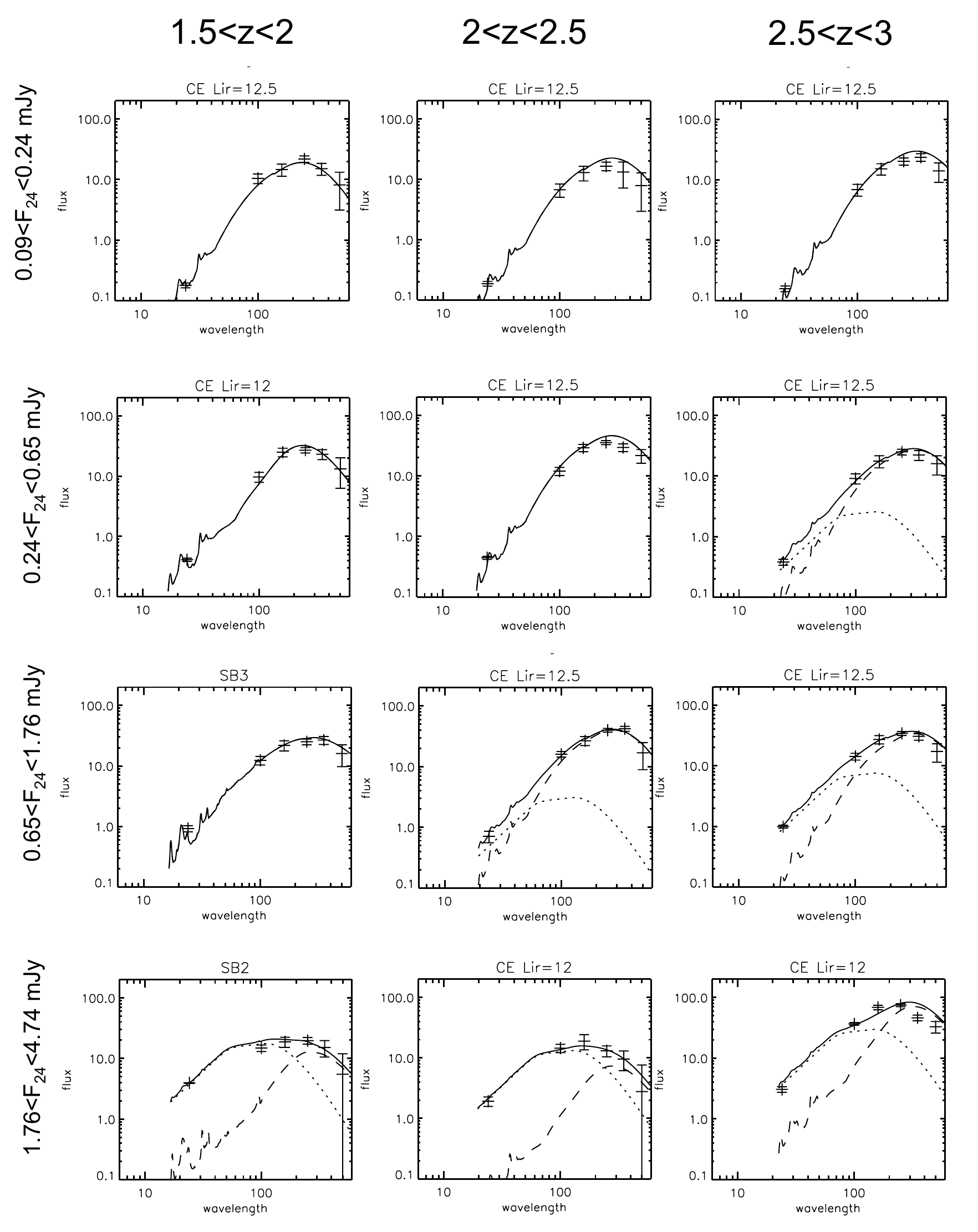}}
 \caption{SEDs of the 12 average templates of DOGs galaxies from our sample. The solid line corresponds to the total SED fit, the dashed line is the host template and the dotted line is the AGN component. The name of the template used is written on each panels. The flux bins are specified on the left side and the redshift bins are written on the top. }
 \label{fig:SED_average}
\end{figure*}

\begin{figure*} 
 \centering
 \resizebox{1.\hsize}{!}{\includegraphics{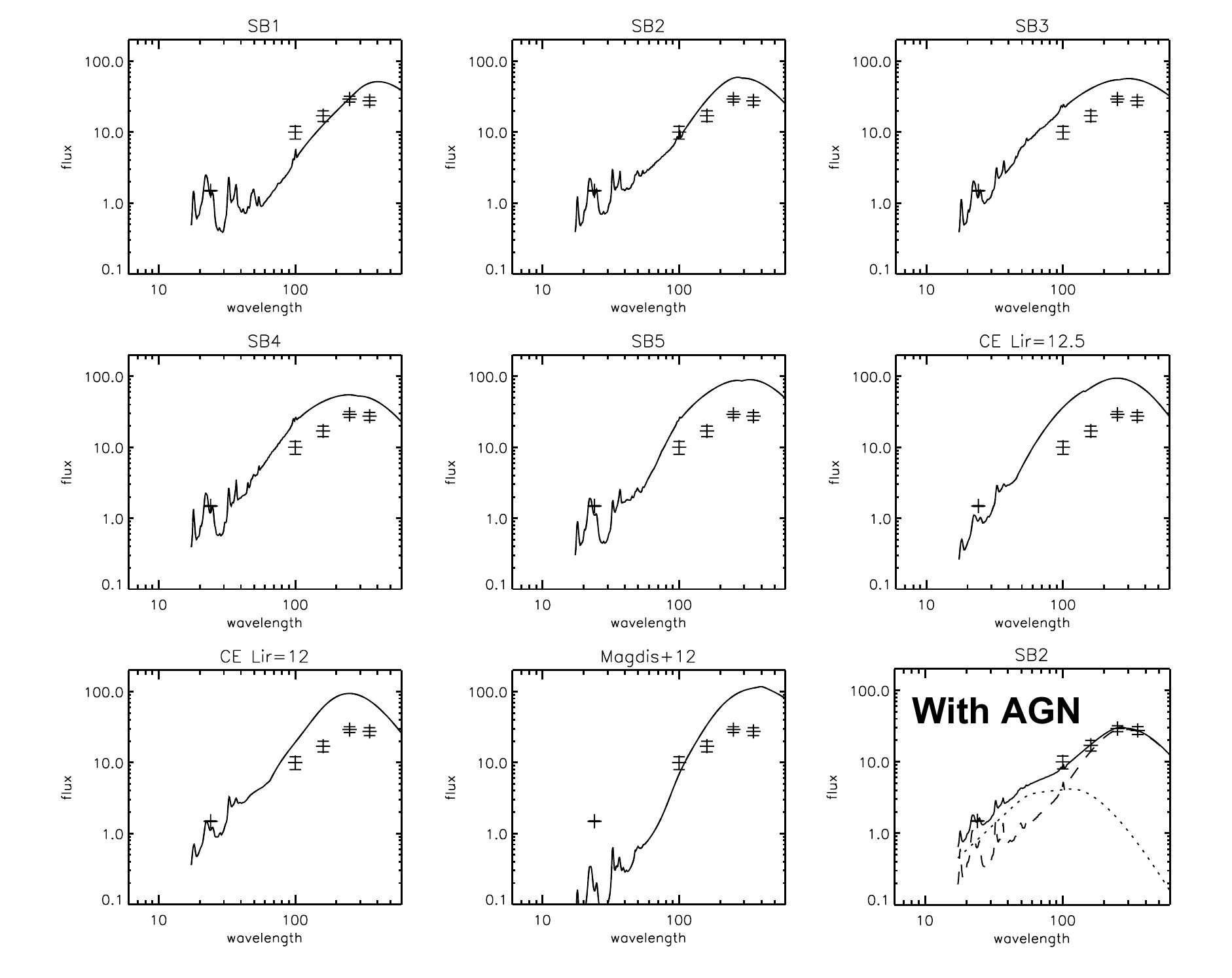}}
\caption{ Following the same format as Fig.\,\ref{fig:SED_average}, this figure shows the results of the SED-fitting procedure for DOG80, one of the 15 DOGs which require an AGN component: the first 7 panels are the results of the SED-fitting with a host component only and the last panel (bottom middle panel) is the acceptable fit, with a contribution of an AGN component ($20\%<$f$_{\rm{AGN}} <$40\%).}
 \label{fig:DOG80}
\end{figure*}

\begin{figure*} 
 \centering
 \resizebox{0.9\hsize}{!}{\includegraphics{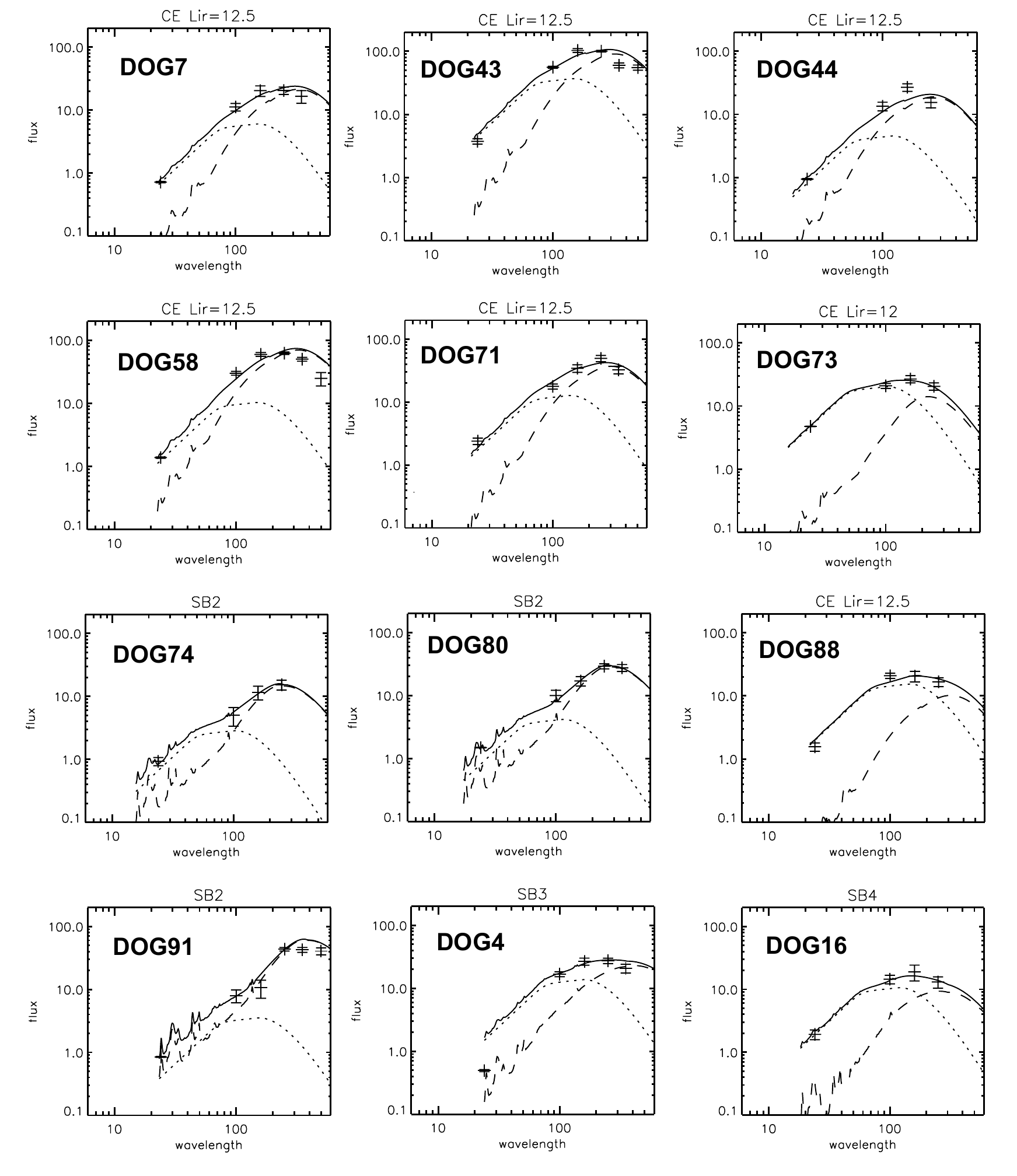}}
 \resizebox{0.9\hsize}{!}{\includegraphics{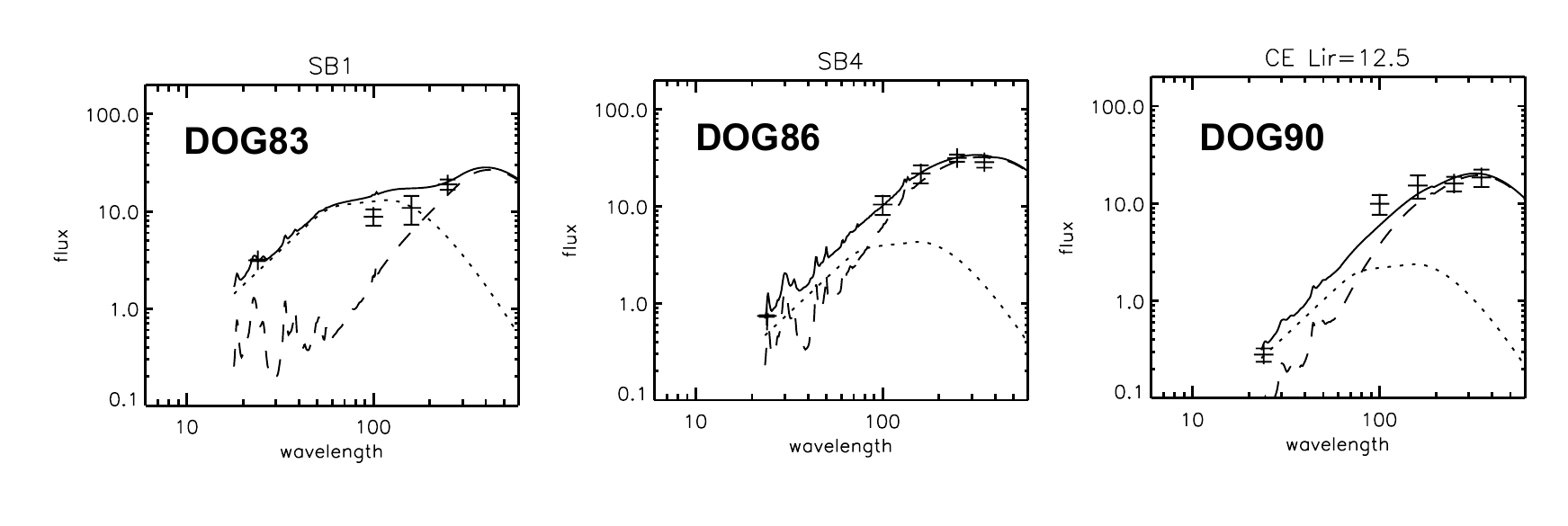}}
 \caption{ Similar to Fig.\,\ref{fig:SED_average}, SEDs of the 15 DOGs which require the contribution of an AGN component.}
 \label{fig:SED_AGN}
\end{figure*}

Due to the expected uncertainties on the AGN fraction for an individual galaxy, we implement the SED-fitting procedure to average DOG SEDs. We divide our DOG sample in 3 different redshift bins (1.5$<$z$<$2.0, 2.0$<$z$<$2.5 and 2.5$<$z$<$3.0) and in four 24~$\mu m$-flux bins (0.09$<$ F$_{24\mu m}  <$0.24, 0.24$<$ F$_{24\mu m}  <$0.65, 0.65$<$ F$_{24\mu m}  <$1.76, and 1.76$<$ F$_{24\mu m}  <$4.74\,mJy). For each redshift bin and each 24~$\mu m$-flux bin, an averaged SED is calculated at 8~$\mu m $, 24~$\mu m $, 100~$\mu m $, 160~$\mu m $, 250~$\mu m $, 350 and 500~$\mu m $, leading to a total of 12 average SEDs. The results of the fits are shown in Fig\,\ref{fig:SED_average} for the average SEDs and in Fig\,\ref{fig:DOG80}\,\&\,\ref{fig:SED_AGN} for the individual sources fitted with an AGN. 

The results from the SED-fitting procedure on the average DOG SEDs are presented on Table\,\ref{tab:SED_fitting}, including the AGN contribution to the total IR flux (f$_{AGN}$), the best-fit template and corresponding $\chi^2$. Our method to get the AGN contribution does not appear to be biased toward one specific template. Independently of the AGN contribution, the two templates that were the most successful at fitting the average SEDs were the CE01 templates for IR luminosities of 10$^{12}$ L$_{\odot}$ (CE12) and 10$^{12.5}$  L$_{\odot}$ (CE12.5). This is as expected, since these templates have PACS/MIR colors consistent with that our DOG sources.

\begin{table*}
\caption{Results from the SED-fitting for the average SEDs per redshift bins and per flux bins. $f_{AGN}$ is the percentage of total 8-1000$\mu$m flux that comes from AGN. We specify the star-forming template best fitting the host galaxy (SB for the starburst templates from \citep{Mullaney:11} and CE 12.5 and CE 12 for the CE01 ULIRG templates.} le
\centering{
\begin{tabular}{|c| c| c| c| c| c| c|} 
\hline
 Redshift  & Flux  & $f_{AGN}$ & error & Template & $\chi^{2}$  &  L$_{ir}$ \\[0.5ex]
   & (mJy)  & \% & \% & (Host galaxy)  &  & L$_{\odot}$
\\ [0.5ex]
\hline\hline 
 1.5$<$z$<$2.0 & 0.09$<$f$<$0.24 & 0 	 & -- & CE 12.5 	& 4.4	   & 9.60e+11 \\
  & 0.24$<$f$<$0.65 & 0	 & -- & CE 12 	& 5.3 &  1.29e+12\\
  & 0.65$<$f$<$1.76 & 0	 & -- & SB3	 	& 5.3 &  1.69e+12\\
  & 1.76$<$f$<$4.74 & 83	 & 3.5 & SB2	 	& 6.5   &  2.42e+12 \\ [1ex]
 \hline
 2$<$z$<$2.5 & 0.09$<$f$<$0.24 & 0  	 & -- & CE 12.5 	& 6.9   &  1.50e+12 \\
  & 0.24$<$f$<$0.65 & 0	 & -- & CE 12.5 	& 34.6   &  3.05e+1\\
  & 0.65$<$f$<$1.76 & 17	 & 5.6 & CE 12.5 	& 0.8    &  5.52e+11\\
  & 1.76$<$f$<$4.74 & 86	 & 15.7 & CE 12 	& 0.8   &    2.43e+12\\ [1ex]
 \hline
 2.5$<$z$<$3.0 & 0.09$<$f$<$0.24 & 0  	 & -- & CE 12.5 	& 21  &  2.42e+12\\
  & 0.24$<$f$<$0.65 & 20	 & 3.5 & CE 12.5 	& 0.4   &  5.60e+11\\
  & 0.65$<$f$<$1.76 & 38	 & 3.2 & CE 12.5 	& 0.1   &  1.69e+12  \\
  & 1.76$<$f$<$4.74 & 59	 & 4.5 & CE 12 	& 13  &  6.68e+12\\ [1ex]
\hline 
\end{tabular}

}
\label{tab:SED_fitting}
\end{table*}

\subsection{Results}

\begin{figure}
 \centering
 \resizebox{1.\hsize}{!}{\includegraphics{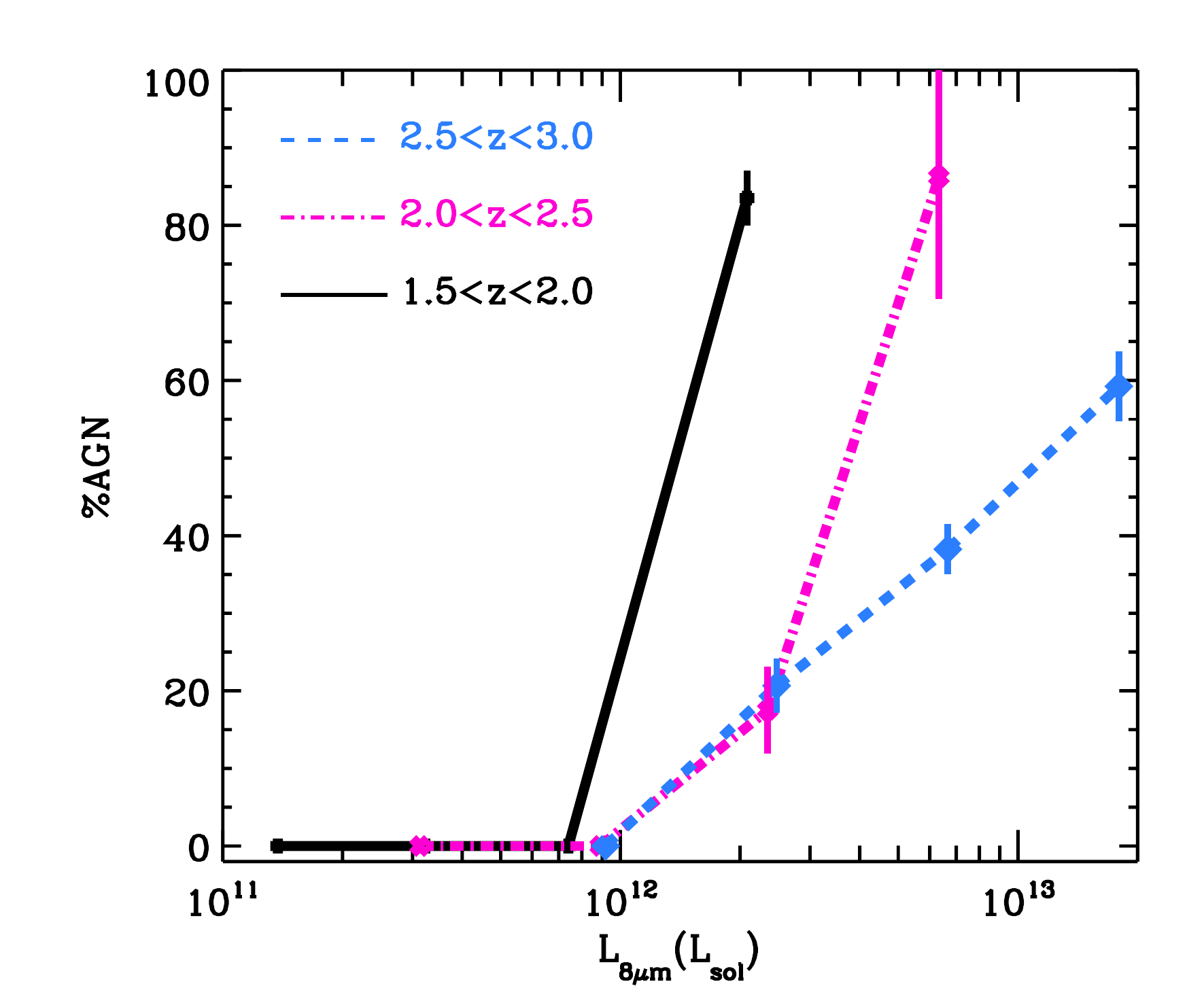}}
  \caption{Evolution of the contribution of the AGN component to the total rest-frame 8-1000$\mu$m flux of the sources as a function of the 8$\mu$m rest-frame luminosity (L$_{8 \mu m}$). The AGN fraction is given for 3 different redshift bins: 1.5$<$z$<$2.0 (solid black line), 2.0$<$z$<$2.5 (pink dot-dashed line) and 2.5$<$z$<$3.0 (bleu dashed line). The AGN contribution is obtained from the fitting procedure described in Sect.~\ref{sec:sed_fitting} on the 12 average SED (see Table\,\ref{tab:SED_fitting}). The general trend is an increasing contribution of the AGN component with respect to the L$_{8 \mu m}$ of the source irrespective of the redshift range.}
 \label{fig:agn_perc_vs_fl24}
\end{figure}

We list the mid-to-far IR photometry and the total IR luminosity of the 95 Herschel-DOGs on Table\,3, indicating also whether an AGN component is included as part of the SED fit. The redshift distribution of our DOG sample peaks
at z$\sim$2, allowing us to use the 24~$\mu$m band as a probe of the mid-IR emission close to the rest frame 8 $\mu$m. This allows us to derive
a L$_{8 \mu m}$ that minimizes the dependence on the choice
of SED template used to perform the $k$-corrections. To determine the rest-frame L$_{8 \mu m}$
we interpolate the CE01 library at the redshift and flux of each
average SED. We are then in a position to study the fractional AGN
contribution to the total 8-1000~$\mu$m output  as a function of the
8$\mu$m rest-frame luminosity (L$_{8 \mu m}$); we do this for the average SEDs at the three redshift bins: $z=1.5-2.0$,
$2.0-2.5$ and $2.5-3.0$ (see Fig.~\ref{fig:agn_perc_vs_fl24}).

We find that the AGN contribution increases globally with increasing
L$_{8 \mu m}$ for all our redshift bins (see Fig.~\ref{fig:agn_perc_vs_fl24}).  This confirms the findings of \citet{Pope:08},
where they report -- based on mid-IR colors of 79 sources within the GOODS field and with 24~$\mu$m fluxes down to 100\,$\mu$Jy -- that low-luminosity DOGs are
primarily powered by star-formation activity. However, only MIPS-70$\mu$m observations were available for their analysis and the inability to sample properly the peak of the SED lead to large uncertainties on the derivation of the dust temperature and the AGN contribution; \citet{Penner:12} extend the study out to far-IR wavebands but miss the faintest DOGs
by focussing on GOODS DOGs with luminosities 10$^{12}$ L$_{\odot} <$ L$_{IR} <$ 10$^{13}$ L$_{\odot}$. Furthermore, \citet{Fiore:08} claimed that even faint DOGs show evidence of hard X-ray emission,
suggesting the presence of an underlying AGN contribution. Within this context our work -- based on the large 2\,deg$^{2}$ area of the COSMOS field and the good sensitivity of
the MIPS-24~$\mu$m observations that allows us to sample a wider range of
24~$\mu$m fluxes (i.e, 80\,$\mu$Jy $< F_{24\mu m} <$ 5 mJy) and the access of far-IR data with {\it Herschel}  -- allows us to conclude
that faint DOGs are mainly star-forming systems while brighter sources
become dominated by an AGN. 

By separating our sample in redshift bins, Fig.~\ref{fig:agn_perc_vs_fl24} also shows that the relation between
AGN fraction and L$_{8 \mu m}$ evolves with redshift:  the slope of the
AGN contribution with respect to the L$_{8 \mu m}$ is steeper at low
redshifts, while at higher redshifts the AGN contribution is less important. This is consistent with the results from \citet{Merloni:08}, where they find that although the accretion rate density onto supermassive black holes (SMBH) and star-formation rate (SFR) densities increase from $z\sim0$ to then decrease beyond z$\sim$2, the decrease in SMBH activity is sharper than that of the SFR. We note that the uncertainties on the derived AGN
contributions are calculated from the formal error output resulting from the
$\chi^{2}$ in the SED-fitting procedure. 

\begin{figure}
 \resizebox{1.1\hsize}{!}{\includegraphics{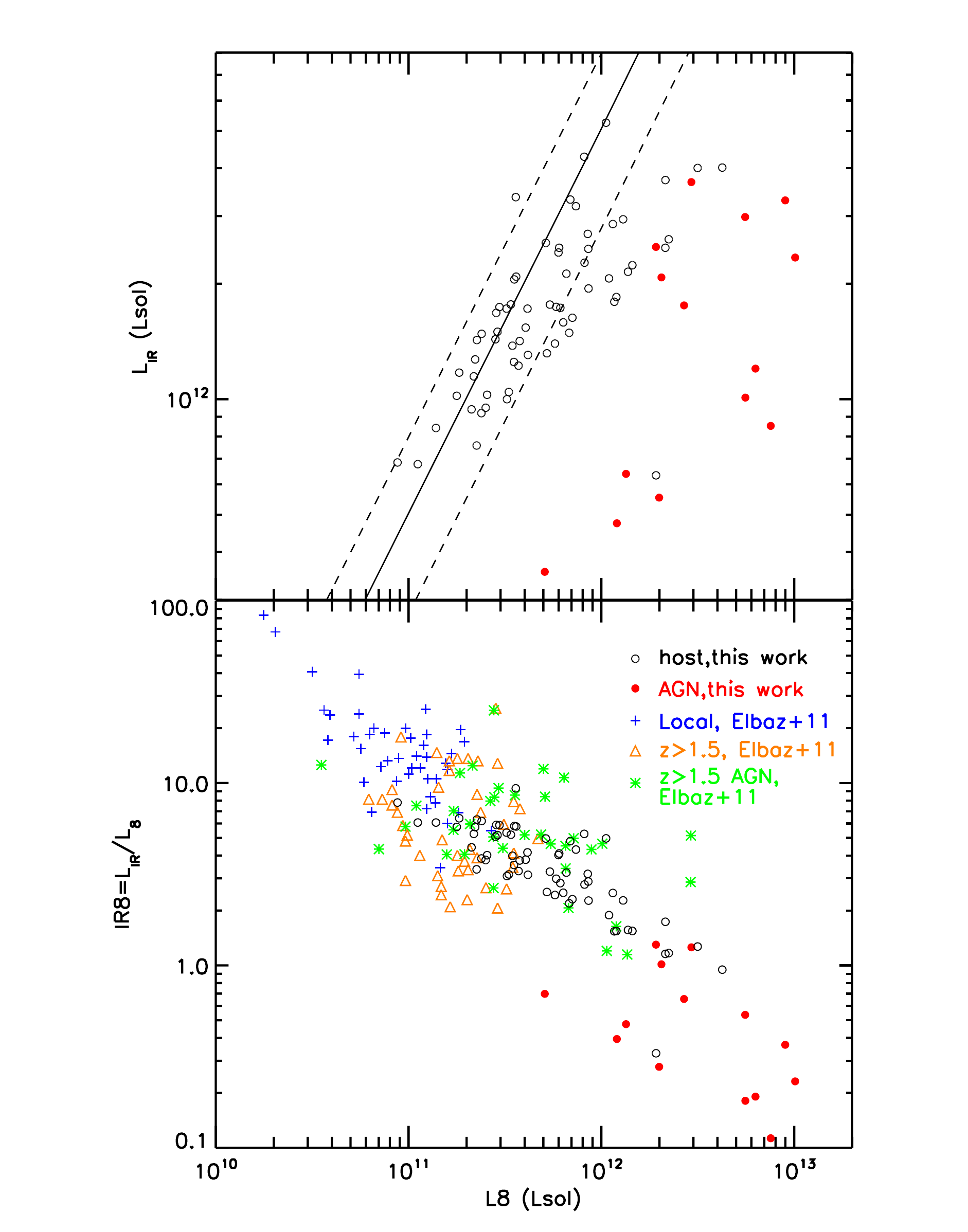}}
\caption{{\it Upper panel:} Comparison of L$_{IR}$ with L$_8$ (rest-frame 8 $\mu$m) for DOGs from our sample (galaxies with host component only are marked with black open circles and AGNs are marked with red filled circles). For comparison, we show the median location of star-forming galaxies from \citet{Elbaz:11} (solid line), with the dashed lines showing the 68\% dispersion. {\it Lower panel:} Variation of the IR8 (=L$_{IR}$/L$_8$) ratio with the 8$\mu$m luminosity for our DOG sample, following the same color code as in the upper panel. For comparison we also plot galaxies from \citet{Elbaz:11}, including local galaxies (blue crosses), star-forming galaxies at z$>$1.5 (orange triangles) and AGNs at z$>$1.5 (green asterisk). See text for details. }
 \label{fig:IR8}
\end{figure}

 In the interest of studying the star formation activity in our sources, we use the results from our SED decomposition to extract the AGN contribution and calculate IR luminosities only due to star formation. The resulting values span the range of 10$^{11}$ L$_{\odot} <$ L$_{IR} <$ 10$^{13}$ L$_{\odot}$, corresponding to one order of magnitude fainter than the analysis by \citet{Penner:12}. We study these IR luminosities as a function of the rest-frame 8$\mu$m luminosity and find that for a given 8$\mu$m luminosity DOG sources whose SEDs are best fit with the addition of an AGN component exhibit significantly lower IR luminosities than DOGs fit with a host-only component (see Fig.\,\ref{fig:IR8}, top panel). In fact, for a given 8$\mu$m luminosity the majority ($\sim$75\%) of DOGs fit by a host-only component display similar IR luminosities to the median star-forming galaxies within the GOODS-Herschel sample from \citet{Elbaz:11}, while AGN-DOGs populate the lower-tail of IR-to-8$\mu$m luminosity ratios (IR8 = L$_{IR}$/L$_{8\mu m}$; see Fig.\,\ref{fig:IR8}, bottom panel). We observe an anti-correlation between the IR8 ratio and the L$_{8\mu m}$ with the AGN-DOGs populating the brightest L$_{8\mu m}$-end.

\begin{figure}
 \centering
 \resizebox{1.\hsize}{!}{\includegraphics{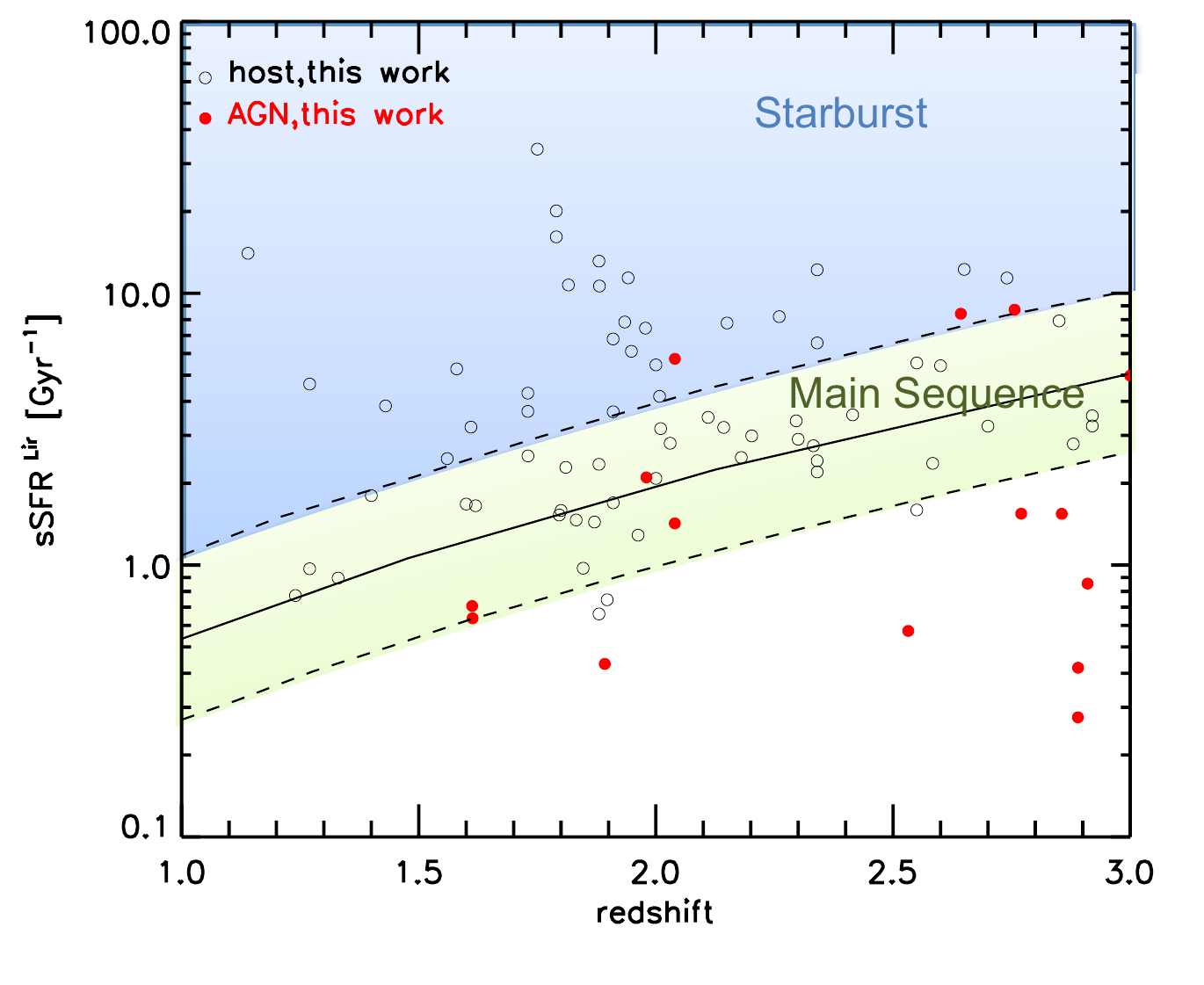}}
 \caption{Redshift evolution of the specific SFR (sSFR = SFR/M$_*$) of DOGs; we distinguish between DOG sources whose SEDs are best fit with a host-only component (black open circles) and with the addition of an AGN component (red filled circles). The solid line represents the star-forming main sequence from \citet{Elbaz:11} and the dashed lines are a factor 2 above and below this fit. See text for details.}
\label{fig:ssfr}
\end{figure}

For each DOG source in our sample we convert the IR luminosity into SFR according to \citet{Kennicutt:98} and adopt the stellar mass from \citet{Ilbert:09}. We consider the redshift evolution of the specific SFR (sSFR = SFR/M$_*$) of our DOG sample and find that the majority of DOGs with no AGN component display sSFRs that place them at or above the main sequence (MS) from \citet{Elbaz:11}, while 50\% of the AGN-DOGs show significantly lower sSFR values (i.e., they lie below the MS)(see Fig.\,\ref{fig:ssfr}). Sources that lie a factor of 2 above the MS are considered as ``starbursts'' by \citet{Elbaz:11}. All but three host-component galaxies lie within a factor of 2 around the MS or in the starburst's zone. The distribution in sSFRs shown in Fig.\,\ref{fig:ssfr} highlights the composite nature of the DOG population: some DOGs are dominated by starburst activity, the majority is undergoing star-formation as part of the MS, while others are dominated by an AGN. This prompts  the idea that DOGs are at the crossroads of the ULIRG-quasar scenario proposed by \citep{Sanders:88a,Sanders:88b,Bussmann:12}, with AGN-DOGs being closer to a quasar phase, where the AGN has already started to quench the star formation (explaining the lower sSFR observed on Fig.\,\ref{fig:ssfr}). 

\subsection{Comparison with IRAC-color AGN selection criteria}

\begin{figure}
 \centering
 \resizebox{1.\hsize}{!}{\includegraphics{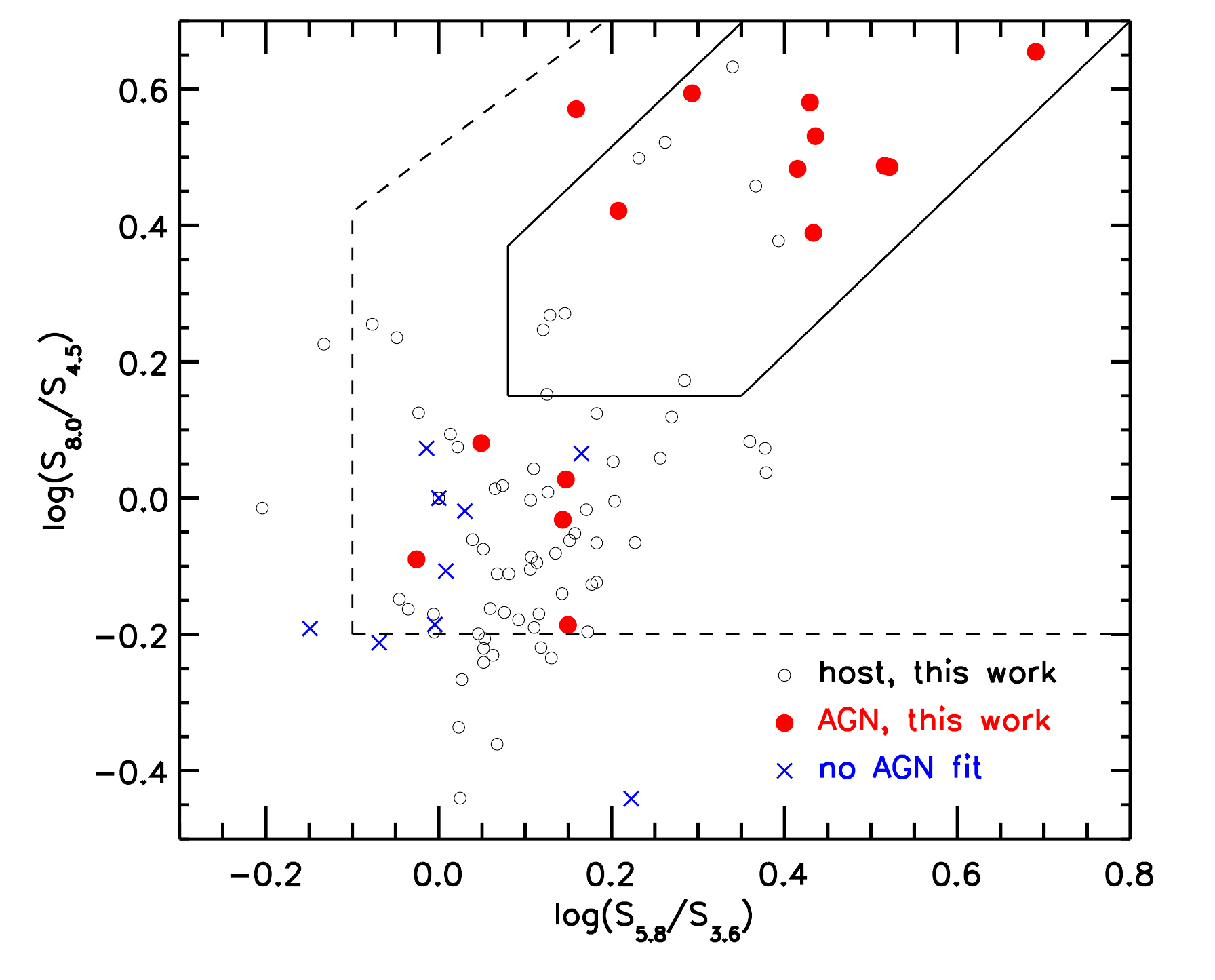}}
  \caption{IRAC colors of host-component DOGs (black open circles) and of the AGN-DOGs (red filled circles) from this work ; we also indicate the IRAC colors of the 9 sources for which no SED fit was possible (blue crosses; see Sect. 4.3 for details). The solid line box is the AGN selection criterion from \citet{Donley:12} and the wider dashed box is from \citet{Lacy:04}.}
 \label{fig:donley_plot}
\end{figure}

 Our SED-fitting analysis identifies 15 Herschel-detected DOGs with an important AGN contribution to the total IR output. We compare our AGN classification of DOGs  to prior approaches relying on an IRAC-color selection.  Fig.\,\ref{fig:donley_plot} shows the IRAC-color selection of AGNs by \citet{Lacy:04}, as well as the refined IRAC-color selection of \citet{Donley:12}, which also includes a power-law criteria in the mid-IR: S$_{3.6} <$ S$_{4.5}$ and S$_{4.5} <$ S$_{5.8}$ and S$_{5.8} <$ S$_{8.0}$. 

The majority of our sources display IRAC colors consistent with the criteria of \citet{Lacy:04}, which would suggest that  90\% of our DOGs are AGNs. However, our SED-fitting analysis indicates that only $\sim$15\% of our sources have a large AGN contribution. Based on this we conclude that relying on the AGN criteria of \citet{Lacy:04} would lead to a lack of precision in selecting AGNs versus galaxies dominated by star formation. On the other hand, more than 50\% of our AGN-DOGs lie within the \citet{Donley:12} criterion, suggesting that it is a more reliable way of selecting AGNs in DOGs when considering merely IRAC colors. However, from the 19 DOGs that lie within the AGN-criteria of \citet{Donley:12} -- and excluding the 3 that do not follow the power-law criteria required by the authors -- only 9 are classified as AGNs following our SED-fitting analysis. That is, 40\% of the Herschel-DOGs with IRAC colors consistent with the criterion of \citet{Donley:12} do not have a significant AGN contribution according to our analysis. Of particular interest is that out of all our AGN-DOGs, six (i.e., $\sim$40\%) are not identified as AGNs based on the criteria by \citet{Donley:12}, four of which do not follow the power-law criterion required. 

On the one hand, our AGN classification is based on the availability of far-IR data for obscured sources such as DOGs. On the other, \citet{Lacy:04} and \citet{Donley:12} classify sources as AGN-dominated based on IRAC-color selections.  When considering these selections side by side, we draw two main conclusions: (1) non PL-DOGs potentially host an AGN that may dominate the far-IR regime even when missed by the IRAC-color selection criteria of \citet{Lacy:04} and \citet{Donley:12}; and (2) PL-DOGs with an AGN according to our SED-fitting procedure can be missed by IRAC colors criteria. We conclude that our method provides an alternate means of determining the composite nature of DOGs.

\section{Dust Temperatures and Masses}
\label{sec:dust}

  It has been well established that interstellar dust absorbs a large fraction of
  the UV/optical radiation from DOGs and reemits it in the IR \citep{Penner:12}
  As such, it is essential that we understand the dust properties of
  these galaxies if we are to understand this potentially-important population of galaxies.  In this section, we derive the
  dust temperatures and masses for our sample of DOGs.  The
  availability of far-IR data from {\it Herschel} is crucial to obtain these
  properties. We are now able to extend previous studies
  on DOGs that did not have access to such high-quality far-IR data
  \citep[e.g.,][]{Dey:08,Bussmann:09,Bussmann:12}. We are also in a position to
  compare results with other recent studies using (limited) {\it Herschel} data on
  DOGs, including that of SPIRE-detected sources (down to only F$_{24\mu m}>$ 0.3\,mJy) with spectroscopic redshifts in the Bootes field by \citet{Melbourne:12} and the study by \citet{Calanog:13} on SPIRE-detected DOGs within COSMOS; no PACS data were available for either study.
  
 \subsection{The Single Temperature Model}
\label{sec:1tmodel}

The availability of {\it Herschel} far-IR data allows us to constrain the peak of the SED in the far-IR regime and
to calculate the dust masses and temperatures of our galaxies with a higher accuracy than previous studies. DecompIR does not provide information on the dust amount of our sources; we fit a blackbody spectrum B$_{\nu}$ of temperature T using far-IR data \citep[see][]{Amblard:14}. To summarize, we perform a
single temperature fit (hereafter 1T model) with an emissivity, $\beta$, of 1.5 to fluxes
long ward of $\lambda_{rest-frame}>$ 40\,$\mu m$. 
The luminosity is then expressed as
\begin{equation}
L(\nu) \propto B(\nu,T_d) \nu^{\beta} .
\end{equation}
Considering $\lambda_{rest-frame}>$ 40\,$\mu m$, we avoid
emission from the AGN that can boost the dust temperature and bias our results
\citep{Netzer:07, Mullaney:11}. For this reason we restrict ourselves to
using only the PACS 100, 160 $\mu m$ and SPIRE 250, 350, 500 $\mu m$
bands for galaxies with z$<$1.7 and only the SPIRE 250, 350,
500 $\mu m$ bands (when available) for z$>$1.7 galaxies.
We require a minimum of 3 data points to fit the SED. At z$<$1.7, 15 DOGs comply with this requirement (from the 16 DOGS at z$<$1.7). At z$>$1.7, from the 24 sources that have SPIRE fluxes above the 3--$\sigma$ limit, 22 have 3 data points.

We enforce
the dust temperature to be constrained between 10 and 95\,K, the
luminosity between 10$^{10}$ and 10$^{14}$\,L$_{\odot}$, acknowledging
the high luminosities of our sample (see Fig.\,\ref{fig:IR8}). We observed the same definition
as \citet{Amblard:14} for the dust luminosity and for the dust mass :

\begin{equation}
L_d(\lambda)=4\pi M_d\kappa(\lambda)B(\lambda,Td)
\end{equation}

\begin{equation}
M_d=L_d/\int{4\pi\kappa(\lambda) B(\lambda,T) d\lambda}
\end{equation}

where $\kappa$ is taken at 850 $\mu$m \citep{Dunne:00} and equal to
0.077 kg$^{-1}$/m$^2$ \citep{Draine:84,Hughes:93}.

\begin{figure}
\centering
\resizebox{1.\hsize}{!}{\includegraphics{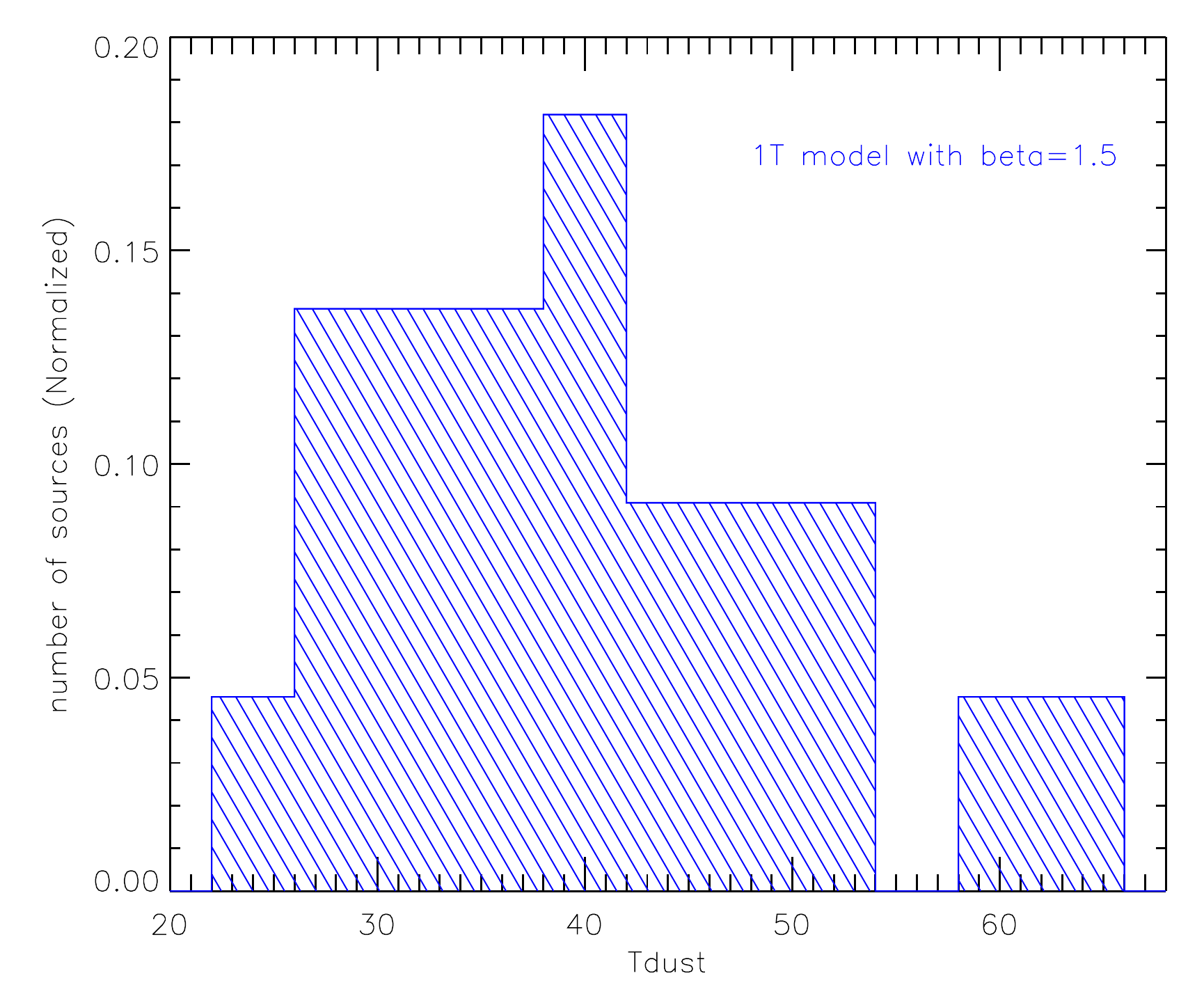}}
 \caption{Distribution of the dust temperature of the 24 DOGs with a detection in the 3 SPIRE bands  with the 1T model (22 sources at z$>1.7$ and 2
sources at lower redshifts). The dust temperature peaks at T$_{d}\sim$40\,K
and is comprised within the range 24$<T_{dust}<$65\,K. }
\label{fig:hist_Tdust}
\end{figure}

We find a median dust temperature of T$_{d}\sim$(40.6\,$\pm$\,9.2)\,K for our sample. 
The dust temperatures of our DOGs are in overall good agreement with
estimates from the literature for other samples of
DOGs. \citet{Bussmann:09} predicted a high dust temperature for DOGs
sources with T$_{d}>$35-60\,K, but this estimate was mainly based on observed
wavelengths shortward of $\sim$350$\mu$m: only 4 of their 12\,DOGs have 350$\mu$m fluxes, the rest of the sample has only upper limits). \citet{Melbourne:12} find lower
dust temperatures for their {\it Herschel}-detected DOG sources (i.e.,
20$<$T$_{d}<$40\,K). They split their sample into bump DOGs and
PL-DOGs and find that the PL-DOGs are less likely to be detected at
far-IR wavelengths using SPIRE than the bump DOGs. They also claim that
SPIRE detections are biased towards very cold sources. We note that our
range of temperature ($\sim$24-65\,K) is wider than that of \citet{Melbourne:12}. Our dust
temperatures are in good agreement with \citet[][within
uncertainties]{Calanog:13} : they find T$_{d}$ = (37\,$\pm$\,6)\,K for detected
PL-DOGs and T$_{d}$ = (35\,$\pm$\,7)\,K for detected bump sources.
As raised in the literature \citep[e.g.,][]{Melbourne:12}, using only SPIRE data tends to underestimate the dust temperature. Therefore, we need to be cautious in our analysis since more than half of our dust temperatures are obtained using only SPIRE data.

\subsection{Discussion}
\label{sec:discussion}

\subsubsection{Effect of the AGN contribution on the dust temperatures}

\begin{figure}
\centering
\resizebox{1.\hsize}{!}{\includegraphics{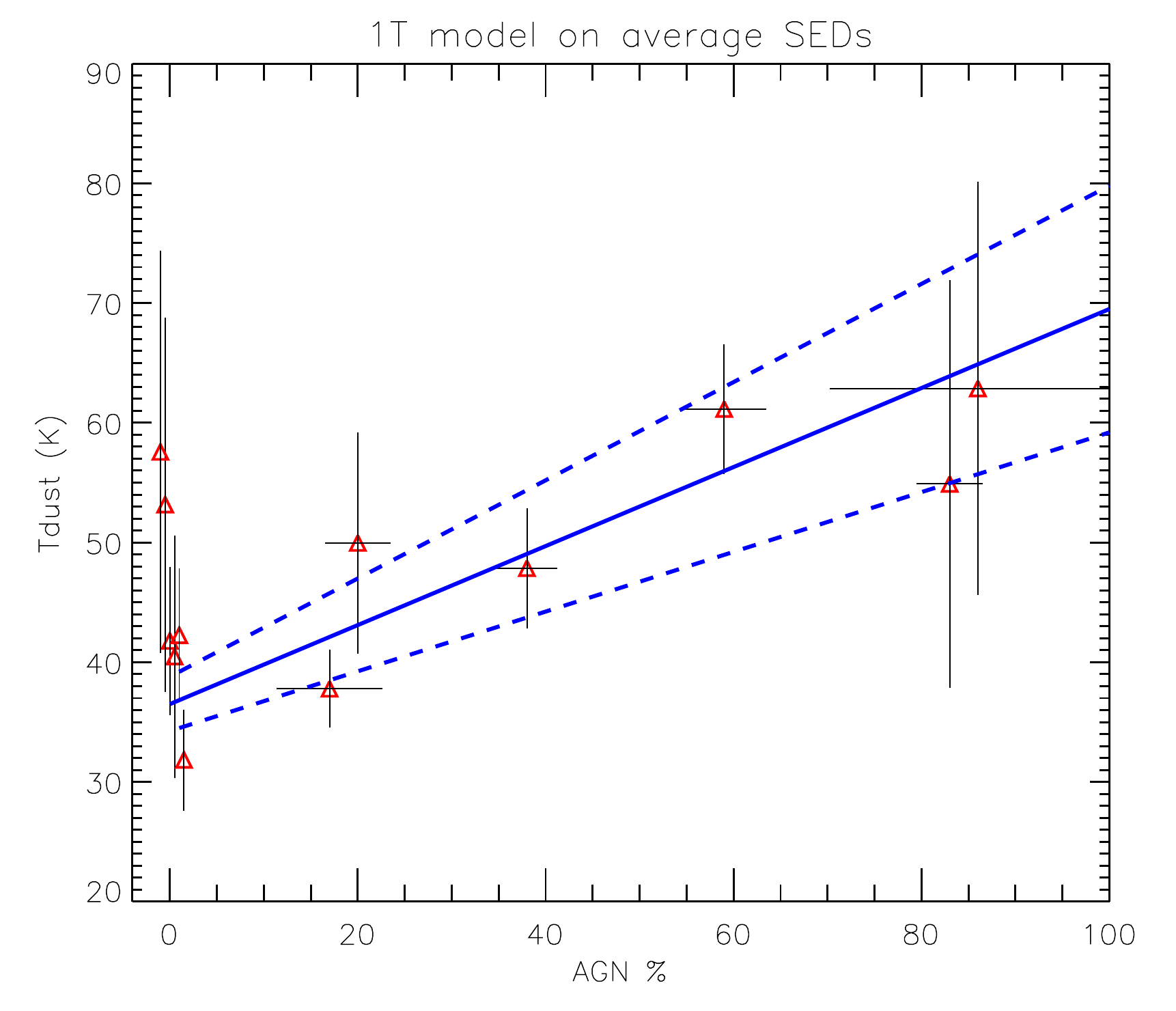}}
 \caption{Evolution of the dust temperature obtained with the 1T model as a function of the AGN fraction. The results are obtained for the 12 average SEDs presented in Table\,\ref{tab:SED_fitting} and detailed in Sec.\,\ref{sec:proc}. To improve clarity, we have slightly changed the x values for the 6 points with no AGN fraction in order to exhibit more clearly the error bars on the figure. The blue line is the best $\chi^{2}$ fit with a slope of 0.33$\pm$0.08 for the 6 average SEDs with an AGN contribution. The dotted lines represent the 1\,$\sigma$ error.}
\label{fig:Tdust_fagn}
\end{figure}

With the goal of improving our accuracy in deriving dust temperature and mass, we use the average SEDs obtained per bin of 24~$\mu$m fluxes and per bin of redshift (see Sec.\,\ref{sec:proc} and in Table\,\ref{tab:SED_fitting} for details) instead of the SEDs of individual DOG sources. The dust temperatures have been derived for the average SEDs following the procedure described in Sec.\,\ref{sec:1tmodel}. These are presented in Fig.\,\ref{fig:Tdust_fagn} as a
function of AGN fractional contribution.  
Roughly half of the average SEDs have no AGN contribution and show a
wide range of T$_{d}$ as seen on the left side of
Fig.\,\ref{fig:Tdust_fagn}.  The two sources with no AGN fraction
 and with the highest dust temperatures show extremely large errors
 bars on the dust temperature (i.e. $\pm$15\,K and $\pm$16\,K) while
 the average dust temperature error for the sample is around 9\,K. 
The rest of the average SEDs have an AGN contribution ranging from
20\% to almost 90\% and are within the same T$_{d}$ range as the
sources with no AGN component. The DOGs' average SEDs with an AGN contribution display a correlation
between the AGN contribution and the dust temperature. We perform a
 best-$\chi^{2}$ fit on the 12 data points, taking into account
the errors both on the x and y axes and find a slope of 0.33$\pm$0.08
with a reduced $\chi^2$ of 0.94. To insure the validity of our fitting
method, we also perform a fit T\,=\,cste with 
\begin{equation*}
cste=\frac{\Sigma\,T/\sigma_T\,^2}{\Sigma\,1/\sigma_T\,^2}=42.6\,K.
\end{equation*}
The reduced $\chi^2$ for this flat fit being 2.53, this gives us a strong indication that our previous fit is valid. We also estimate the Spearman's (rho) rank correlation of T$_{dust}$ and the AGN percentage. The significance is low (0.04),  which indicates a significant correlation.
The correlation between the dust temperature and the AGN percentage for the average SEDs with AGN contribution is thus real, even though the slope is small. However, we do not confirm the presence of a general trend between the AGN fraction and the dust temperature of the sources since half of the data points with no AGN activity present similar dust temperatures than the average SEDs with a large contribution from an AGN.

\subsubsection{Effect of the AGN contribution on the dust masses}

In addition to dust temperatures, the 1T model fitting procedure also
provides us with the dust masses of our sample sources. We obtain a range for the entire sample of 7 $\times 10^{7}<M_{dust}<
10^{9}$\,M$_{\odot}$ and a 
median dust mass of $\sim$\,(3\,$\pm$\,3) $\times$
10$^{8}$\,M$_{\odot}$. 
  Our
results are in very good agreement with \citet{Bussmann:09} who found
a median dust mass of 3$\times$10$^{8}$\,M$_{\odot}$ for their sample of 31 of the brightest DOGs (F$_{24\mu m} > $0.8\,mJy) in the Bootes Field and with HST imaging. Their sample is dominated by sources with a power law in the mid-IR IRAC bands, which is a signature of the presence of an AGN \citep{Donley:07}.

DOGs are believed to be an intermediate AGN phase between high-redshift sub millimeter galaxies (SMGs) and quasars at z$\sim$2 \citep{Bussmann:12}. 
Accounting for uncertainty in $\kappa$ which could be as much as a factor of three, the median dust mass of our sample is not different from those estimated for high-redshift SMGs by \citet{Magnelli:12} (M$_{d} \sim$ 10$^{9}$\,M$_{\odot}$). \citet{Pope:08} found that 30\,\% of the SMGs from their sample also satisfy the DOG criteria, and of those SMG-DOGs, 30\,\% are AGN dominated.
DOGs could then be the descendants of these SMGs with similar dust content, but representing a more advanced AGN-phase than could later quench the star-formation and lead to elliptical galaxies.

\section{Summary and conclusions}
\label{sec:ccl}

We carry out a study that aims to understand the composite nature of 24$\mu m$-bright Dust-Obscured Galaxies (DOGs).
These sources are a subset of ULIRGs at high redshift ($z\sim2$) with F$_{24\mu m}$/F$_{R}>$ 982 . ULIRGs are considered to represent an important phase in the
evolution of galaxies as they are linked to the formation of massive galaxies via gas-rich starbursting mergers followed by an AGN-driven quenching of the star-formation
\citep[e.g.,][]{Sanders:88a,Sanders:88b}. Recent studies
\citep{Dey:08,Bussmann:09,Bussmann:12} have suggested a similar
evolutionary sequence where DOGs are an important intermediate phase between gas-rich
major mergers (traced by submillimeter galaxies, SMGs) and quasars at z$\sim$2. These studies describe an
evolutionary scenario in which the starbursting nature of SMGs evolves into the composite nature of
DOGs as an underlying AGN grows; this is followed by a quasar phase that terminates star formation, leading to the formation of a passive, massive elliptical
galaxy. Within this context, DOGs could provide a key insight to an extremely dusty stage in the evolution of
galaxies at $z\sim2$, where both AGN and star formation activity coexist. Their composite nature was until relatively recently inaccessible prior to the availability of sensitive
mid- to far-infrared data. 

We base our work on a sample of 95 {\it Herschel}-detected DOG sources. We perform SED-fitting on our sources using composite spectra to obtain AGN contributions, dust temperatures and dust masses. We summarize below our results and our conclusions:

\begin{enumerate}

\item  DOGs with the brightest 24$\mu m$ fluxes (F$_{24\mu m} > 1$\,mJy) present significantly bluer PACS/24$\mu m$ colors than other 24$\mu m$-selected sources. These bluer colors may be explained by templates containing an AGN contribution of at least 25\%.

\item Among our sample of 95 sources, 74\% are fit by a host galaxy template while for 16\% require an additional AGN component. The remaining 10\% of the sample could not be properly fit, likely due to inaccurate photometric redshifts.

\item  Faint DOG sources with L$_{8\mu m}<$ 10$^{12}\,L_{\odot}$ are dominated by star-formation at all redshifts, while DOGs brighter than L$_{8\mu m}>$ 2$\times$10$^{12}$ L$_{\odot}$ display a high contribution ($>$20\%) from an AGN component. 

\item  DOGs with no significant AGN contribution are mainly located within the star-forming main sequence as defined in \citet{Elbaz:11}. Those identified as AGN-DOGs present the lowest IR8 (=L$_{IR}$/L8) ratio of our sample and 50\% of them lie below this sequence, with significantly lower specific star-formation rates. This results support the evolutionary scenario where DOGs may represent a  transition phase between high-redshift starburst-dominated SMGs and red-dead ellipticals, passing through an AGN-phase that would quench star formation.

\item  The dust temperature of DOGs peaks at (40\,$\pm$\,9)\,K and our range of temperatures (24$<T_{d}<$65\,K) is overall in good agreement with the literature \citep{Bussmann:09,Bussmann:12,Melbourne:12,Calanog:13}. DOGs with a contribution from an AGN in the far-IR of at least 60\,\% have dust temperatures $>$50\,K, suggesting that the AGN heats the dust of its host galaxy.
We find a median dust mass of $\sim$ (3\,$\pm$\,3) $\times$ 10$^{8}$\,M$_{\odot}$ for our sample consistent previous analysis in the literature \citep{Bussmann:12}. 

\end{enumerate}

This work sheds light on DOG sources and their underlying composite nature, bringing unequivocally to light that mid-IR bright DOGs are powered by an AGN. The submillimeters facilities in the near future, such as CCAT and ALMA will provide critical insight to study the AGN properties of these obscured ULIRGs at z$\sim$2.

This paper is the first of a series on the panchromatic view of DOGs. In this paper we focus on their far-IR properties, while in the upcoming papers we will focus on their X-ray properties based on X-ray stacking analysis and on their contribution to the Cosmic X-ray Background. The far-IR/radio correlation of these sources and their radio properties using JVLA-COSMOS observations presented in \citet{Smolcic:14} will be detailed in a forthcoming paper.

\bigskip

{\it Acknowledgments:}  

COSMOS is based on observations with the NASA/ESA Hubble Space Telescope, obtained at the Space Telescope Science Institute, which is operated by AURA, Inc., under NASA contract NAS 5-26555; also based on data collected at: the Subaru Telescope, which is operated by the National Astronomical Observatory of Japan; XMM-Newton, an ESA science mission with instruments and contributions directly funded by ESA Member States and NASA; the European Southern Observatory, Chile; Kitt Peak National Observatory, Cerro Tololo Inter-American Observatory, and the National Optical Astronomy Observatory, which are operated by the Association of Universities for Research in Astronomy (AURA), Inc., under cooperative agreement with the National Science Foundation; the National Radio Astronomy Observatory, which is a facility of the National Science Foundation operated under cooperative agreement by Associated Universities,Inc; and the Canada-France-Hawaii Telescope, operated by the National Research Council of Canada, the Centre National de la Recherche Scientifique de France, and the University of Hawaii. \\
PACS has been developed by a consortium of institutes led by MPE (Germany) and including UVIE (Austria); KU Leuven, CSL, IMEC (Belgium); CEA, LAM (France); MPIA (Germany); INAF-IFSI/OAA/OAP/OAT, LENS, SISSA (Italy); IAC (Spain). This development has been supported by the funding agencies BMVIT (Austria), ESA-PRODEX (Belgium), CEA/CNES (France), DLR (Germany), ASI/INAF (Italy), and CICYT/MCYT (Spain). \\
SPIRE has been developed by a consortium of institutes led by Cardiff University (UK) and including University of Lethbridge (Canada), NAOC (China), CEA, LAM (France), IFSI, University of Padua (Italy), IAC (Spain), Stockholm Observatory (Sweden), Imperial College London, RAL, UCL-MSSL, UKATC, University of Sussex (UK), Caltech, JPL, NHSC, University of Colorado (USA). This development has been supported by national funding agencies: CSA (Canada); NAOC (China); CEA, CNES, CNRS (France); ASI (Italy); MCINN (Spain); SNSB (Sweden); STFC, UKSA (UK) and NASA (USA).
SPIRE has been developed by a consortium of institutes led by Cardiff Univ. (UK) and including Univ. Lethbridge (Canada); NAOC (China); CEA, LAM (France); IFSI, Univ. Padua (Italy); IAC (Spain); Stockholm Observatory (Sweden); Imperial College London, RAL, UCL-MSSL, UKATC, Univ. Sussex (UK); Caltech, JPL, NHSC, Univ. Colorado (USA). This development has been supported by national funding agencies: CSA (Canada); NAOC (China); CEA, CNES, CNRS (France); ASI (Italy); MCINN (Spain); SNSB (Sweden); STFC, UKSA (UK); and NASA (USA).

\begin{table*}
\caption{DOGs sample}
\begin{tabular}{|c|c|c|c|c|c|c|c|c|c|c|}
 DOG ID &  redshift &  F$_{24}$ &  F$_{100}$ &  F$_{160}$ & F$_{250}$ & F$_{350}$ & F$_{500}$ & flag AGN & $\chi^2$ &  L$_{ir}$  \\
 	      &		       &   (mJy)	  &   (mJy)	       &   (mJy)	    &   (mJy)      &   (mJy)	    &   (mJy)	&                &                &   (L$_{\odot}$)    \\
 \hline \hline
       0   &   1.87   &   0.386  &    9.798   &    14.33   &    35.05  &     17.40  &     14.67  &     0  &    2.90 &  1.02e+12		\\
       1   &   1.60   &   0.178  &    11.48   &    10.53   &    23.54  &     16.14  &     10.17  &     0  &    9.59 &  8.41e+11		\\
       2   &   2.65   &   0.176  &    9.251   &    11.81   &    21.27  &     19.14  &     14.26  &     0  &    6.22 &  3.41e+12		\\
       3   &   2.34   &   0.491  &    16.73   &    26.67   &    27.36  &     20.72  &     7.077  &     0  &    3.00 &  2.48e+12		\\
       4   &   3.00   &   0.097  &    9.961   &    27.31   &    25.93  &     14.49  &     3.826  &     3  &   0.409 &  3.67e+12		\\
       5   &   1.00   &   0.404  &    9.088   &    20.96   &    12.49  &     8.223  &     6.096  &     0  &    37.8 &  1.30e+12		\\
       6   &   1.40   &   0.440  &    7.806   &    16.88   &    15.89  &     13.25  &     9.133  &     0  &    20.5 &  6.84e+11		\\
       7   &   2.85   &   0.326  &    8.145   &    13.69   &    22.02  &     18.26  &     5.261  &     2  &    2.19 &  1.20e+12		\\
       8   &   2.34   &   0.402  &    8.467   &    24.79   &    22.32  &     20.23  &     9.081  &     0  &    3.00 &  2.48e+12		\\
       9   &   1.89   &   0.312  &    15.11   &    10.92   &    10.72  &     10.60  &     8.063  &     0  &    7.78 &  1.49e+12		\\
      10   &   1.14   &   0.103  &    5.576   &    16.27   &    15.17  &     17.43  &     5.545  &     0  &    3.56 &  4.15e+11		\\
      11   &   1.98   &   0.156  &    14.28   &    18.67   &    25.49  &     17.19  &     2.676  &   -99  &   -99   &   -99			\\
      12   &   1.88   &   0.236  &    11.65   &    29.26   &    19.52  &     7.441  &     3.868  &     0  &    24.5 &  1.587e+12		\\
      13   &   1.81   &   0.444  &    14.62   &    43.96   &    31.01  &     17.75  &     7.619  &     0  &  0.0739 &  9.502e+11		\\
      14   &   1.79   &   0.201  &    6.918   &    32.55   &    30.33  &     19.84  &     9.915  &     0  &    31.7 &  2.460e+12		\\
      15   &   2.30   &   0.326  &    8.989   &    18.93   &    30.08  &     21.45  &     20.49  &     0  &    8.01 &  2.068e+12		\\
      16   &   2.04   &   0.656  &    9.732   &    20.43   &    30.88  &     34.16  &     15.04  &     4  &   0.668 &  1.753e+12		\\
      17   &   2.34   &   0.716  &    11.18   &    20.34   &    20.45  &     16.57  &     8.984  &     0  &   0.311 &  1.760e+12		\\
      18   &   2.88   &   0.330  &    17.99   &    54.18   &    43.51  &     35.79  &     32.05  &     0  &    12.6 &  4.012e+12		\\
      19   &   2.74   &   0.223  &    12.73   &    22.93   &    32.10  &     20.40  &     5.487  &     0  &   0.509 &  4.354e+12		\\
      20   &   2.41   &   0.665  &    16.44   &    27.88   &    17.78  &     14.48  &     0.769  &     0  &    2.57 &  1.842e+12		\\
      21   &   1.90   &   0.363  &    6.506   &    9.178   &    21.98  &     19.80  &     13.77  &     0  &    4.31 &  1.538e+12		\\
      22   &   2.15   &   0.319  &    9.573   &    31.35   &    50.37  &     43.09  &     29.44  &     0  &    29.9 &  1.798e+12		\\
      23   &   2.11   &   0.400  &    9.302   &    26.70   &    34.57  &     26.41  &     13.75  &     0  &    15.8 &  2.122e+12		\\
      24   &   2.14   &   0.208  &    5.767   &    15.81   &    18.79  &     12.01  &     5.824  &     0  &    32.7 &  2.412e+12		\\
      25   &   2.50   &   0.414  &    4.650   &    14.08   &    35.78  &     28.10  &     10.19  &   -99  &   -99   &   -99		\\
      26   &   2.29   &   0.485  &    7.434   &    15.55   &    26.60  &     25.71  &     25.31  &     0  &    9.11 &  3.31e+12		\\
      27   &   1.84   &   0.285  &    7.054   &    23.17   &    24.96  &     21.93  &     9.473  &     0  &    5.71 &  9.19e+11		\\
      28   &   1.79   &   0.400  &    7.824   &    16.42   &    12.27  &     14.91  &     6.783  &     0  &    3.61 &  1.59e+12		\\
      29   &   1.80   &   0.532  &    7.399   &    21.88   &    28.97  &     24.07  &     13.26  &   -99  &   -99   &   -99		\\
      30   &   1.91   &   0.287  &    8.802   &    23.61   &    27.99  &     26.50  &     17.52  &     0  &    2.37 &  1.14e+12		\\
      31   &   1.80   &   0.601  &    10.28   &    32.24   &    50.22  &     57.26  &     37.58  &     0  &   0.503 &  1.67e+12		\\
      32   &   1.43   &   0.271  &    27.75   &    50.52   &    54.32  &     22.67  &     22.84  &     0  &    47.8 &  1.42e+12   		\\
      33   &   1.24   &   0.405  &    7.274   &    12.76   &    24.40  &     25.07  &     12.53  &     0  &  0.0981 &  6.33e+11		\\
      34   &   1.73   &   0.090  &    3.589   &    13.36   &    18.04  &     8.671  &     1.258  &     0  &    1.04 &  1.04e+12		\\
      35   &   1.83   &   0.669  &    7.372   &    18.38   &    29.18  &     21.97  &     15.27  &     0  &    3.06 &  1.73e+12		\\
      36   &   1.88   &   0.388  &    12.93   &    17.25   &    11.61  &     5.563  &     3.736  &     0  &    13.3 &  1.63e+12		\\
      37   &   2.18   &   0.185  &    11.57   &    16.36   &    20.27  &     11.07  &     7.944  &     0  &    4.96 &  2.85e+12		\\
      38   &   1.68   &   0.846  &    7.959   &    10.58   &    43.39  &     42.47  &     40.47  &     0  &    2.16 &  9.40e+11		\\
      39   &   1.56   &   0.280  &    9.949   &    15.29   &    16.03  &     18.52  &     12.97  &     0  &    10.0 &  1.25e+12		\\
      40   &   1.71   &   0.412  &    17.15   &    40.69   &    38.81  &     23.32  &     9.754  &     0  &    4.66 &  1.02e+12		\\
      41   &   2.00   &   0.737  &    10.41   &    21.83   &    31.49  &     28.50  &     17.59  &     0  &    9.79 &  1.17e+12		\\
      42   &   1.79   &   0.282  &    5.335   &    14.58   &    22.06  &     15.15  &     3.131  &     0  &    8.24 &  1.37e+12		\\
      43   &   2.75   &   0.277  &    7.221   &    21.95   &    14.15  &     16.22  &     13.70  &     3  &    15.7 &  8.33e+12		\\
      44   &   2.03   &   1.038  &    29.37   &    43.28   &    34.96  &     28.66  &     7.363  &     2  &    13.4 &  4.74e+11		\\
      45   &   1.91   &   0.248  &    5.540   &    14.59   &    26.78  &     19.12  &     7.714  &     0  &    5.65 &  1.43e+12		\\
      46   &   1.81   &   0.551  &    16.28   &    28.14   &    17.47  &     10.42  &     6.205  &     0  &    4.42 &  1.72e+12		\\
      47   &   2.12   &   0.179  &    5.090   &    8.722   &    25.19  &     24.05  &     19.96  &   -99  &   -99   &   -99		\\
      48   &   2.35   &   0.386  &    9.840   &    30.80   &    33.42  &     44.45  &     28.70  &     0  &    11.8 &  4.28e+12		\\
      49   &   1.96   &   0.130  &    6.263   &    12.26   &    20.61  &     23.57  &     12.48  &     0  &    1.40 &  1.48e+12		\\
      50   &   2.92   &   0.296  &    9.159   &    25.80   &    40.72  &     28.25  &     21.91  &     0  &    8.62 &  3.72e+12		\\
      51   &   2.55   &   0.559  &    9.291   &    20.90   &    20.95  &     4.990  &     16.31  &     0  &    1.26 &  1.82e+12		\\
      52   &   1.94   &   0.208  &    6.285   &    11.61   &    19.85  &     18.14  &     7.037  &     0  &    2.37 &  1.72e+12		\\
      53   &   2.26   &   0.379  &    9.616   &    28.90   &    38.44  &     34.29  &     20.98  &     0  &    8.56 &  1.51e+12		\\
      54   &   1.88   &   0.704  &    4.882   &    29.71   &    30.02  &     36.79  &     7.011  &     0  &   0.672 &  1.41e+12		\\
      55   &   1.61   &   0.187  &    7.526   &    18.13   &    19.77  &     23.50  &     15.58  &     0  &    25.1 &  2.08e+12		\\
      56   &   1.61   &   0.284  &    7.408   &    19.38   &    21.56  &     17.52  &     9.390  &     0  &    2.67 &  1.00e+12		\\
      57   &   1.73   &   0.537  &    5.251   &    16.61   &    29.93  &     25.76  &     9.697  &     0  &    24.1 &  1.73e+12		\\
      58   &   2.77   &   0.165  &    7.150   &    14.23   &    13.25  &     8.367  &     8.701  &     2  &    20.9 &  2.33e+12		\\
      59   &   1.94   &   0.330  &    14.65   &    24.56   &    26.93  &     21.13  &     10.50  &     0  &   0.369 &  1.94e+12		\\
      60   &   1.27   &   0.433  &    10.52   &    27.31   &    44.89  &     44.77  &     34.52  &     0  &    9.25 &  7.57e+11 \\
\end{tabular}

\medskip
{\it flag AGN} is the contribution from an AGN to the host galaxy obtained from DecompIR: (0): only host galaxy, (1): \% AGN $\le$ 20\%,  (2): 20 $<$ \% AGN $\le$ 40\%,  (3): 40 $<$ \% AGN $\le$ 70\%,  (4): \% AGN $>$ 70\%
\end{table*}

\begin{table*}
      \contcaption{}
\begin{tabular}{|c|c|c|c|c|c|c|c|c|c|c|}
 DOG ID &  redshift &  F$_{24}$ &  F$_{100}$ &  F$_{160}$ & F$_{250}$ & F$_{350}$ & F$_{500}$ & flag AGN & $\chi^2$ &  L$_{ir}$  \\
 	      &		       &   (mJy)	  &   (mJy)	       &   (mJy)	    &   (mJy)      &   (mJy)	    &   (mJy)	&                &                &   (L$_{\odot}$)    \\
 \hline \hline
      61  &    2.00  &    0.745   &   15.31   &    25.86  &     38.35   &    47.96   &    29.10  &     0  &    50.1 &  3.18e+12		\\
      62  &    1.58  &    0.505   &   6.231   &    31.96  &     8.530   &    45.61   &    8.272  &     0  &    7.96 &  1.22e+12		\\
      63  &    1.88  &    0.621   &   17.10   &    39.68  &     41.01   &    29.16   &    21.66  &     0  &    6.60 &  6.77e+11		\\
      64  &    2.55  &    0.366   &   8.821   &    15.62  &     8.706   &    3.102   &    6.735  &     0  &    2.24 &  2.61e+12		\\
      65  &    1.76  &    0.419   &   9.343   &    19.55  &     28.59   &    27.63   &    20.37  &   -99  &   -99   &   -99				\\
      66  &    2.03  &    0.407   &   14.68   &    28.11  &     22.86   &    15.96   &    10.10  &     0  &    2.44 &  2.69e+12		\\
      67  &    2.85  &    0.497   &   16.95   &    27.73  &     10.42   &    8.087   &    11.99  &     0  &    14.7 &  4.00e+12		\\
      68  &    1.94  &    0.943   &   13.47   &    26.86  &     15.43   &    7.120   &    2.860  &   -99  &   -99   &   -99		\\
      69  &    1.62  &    0.925   &   43.88   &    66.54  &     54.21   &    32.34   &    6.816  &   -99  &   -99   &   -99				\\
      70  &    1.93  &    0.437   &   7.177   &    19.51  &     21.00   &    8.710   &    8.849  &     0  &    1.69 &  1.39e+12		\\
      71  &    2.53  &    0.132   &   16.52   &    44.25  &     51.89   &    34.34   &    23.83  &     3  &    1.20 &  2.98e+12		\\
      72  &    2.91  &    0.555   &   7.702   &    23.22  &     19.01   &    19.22   &    10.53  &   -99  &   -99   &   -99		\\
      73  &    1.61  &    0.395   &   13.92   &    23.40  &     31.45   &    13.30   &    11.82  &     4  &    1.44 &  2.49e+12		\\
      74  &    1.61  &    0.266   &   7.749   &    20.32  &     20.94   &    10.23   &    1.243  &     3  &   0.200 &  3.54e+11		\\
      75  &    2.70  &    0.256   &   6.372   &    13.61  &     29.14   &    24.31   &    23.73  &     0  &    1.67 &  3.36e+12		\\
      76  &    2.92  &    0.465   &   13.72   &    41.37  &     39.27   &    37.63   &    28.69  &     0  &    1.60 &  2.94e+12		\\
      77  &    2.00  &    1.487   &   10.00   &    17.00  &     29.15   &    27.47   &    5.519  &     0  &    14.1 &  1.48e+12		\\
      78  &    1.98  &    0.554   &   22.49   &    71.89  &     74.66   &    50.33   &    50.14  &   -99  &   -99   &   -99		\\
      79  &    2.11  &    0.359   &   14.87   &    21.01  &     25.52   &    18.83   &    15.39  &     0  &    11.0 &  2.55e+12		\\
      80  &    1.89  &    4.742   &   20.84   &    26.75  &     20.47   &    16.42   &    8.412  &     2  &   0.761 &  6.38e+11		\\
      81  &    2.33  &    1.385   &   30.01   &    59.63  &     62.33   &    50.04   &    24.83  &     0  &    3.97 &  1.31e+12		\\
      82  &    1.97  &    0.625   &   12.46   &    20.73  &     26.62   &    27.96   &    11.84  &     0  &    5.23 &  1.26e+12		\\
      83  &    1.98  &    1.059   &   28.67   &    36.87  &     32.63   &    40.06   &    27.67  &     3  &    15.6 &  2.07e+12		\\
      84  &    1.91  &    0.578   &   7.835   &    13.90  &     48.09   &    51.41   &    32.16  &     0  &    3.45 &  1.74e+12		\\
      85  &    2.58  &    1.917   &   14.49   &    18.95  &     13.09   &    9.595   &    2.781  &     0  &    15.9 &  2.26e+12		\\
      86  &    2.91  &    0.759   &   11.60   &    21.18  &     35.11   &    49.91   &    24.43  &     2  &   0.021 &  1.00e+12		\\
      87  &    2.34  &    0.637   &   7.290   &    17.49  &     20.34   &    23.81   &    13.98  &     0  &    6.19 &  2.14e+12		\\
      88  &    2.64  &    3.744   &   55.74   &    102.9  &     100.3   &    59.92   &    55.24  &     1  &    8.16 &  3.29e+12		\\
      89  &    2.20  &    0.860   &   12.87   &    22.32  &     12.75   &    19.16   &    21.22  &     0  &    19.6 &  5.25e+12		\\
      90  &    2.89  &    0.671   &   16.75   &    27.59  &     41.20   &    36.35   &    4.689  &     1  &    4.33 &  5.54e+11		\\
      91  &    2.88  &    0.931   &   5.011   &    11.63  &     15.30   &    13.39   &    16.10  &     1  &    5.19 &  8.52e+11		\\
      92  &    2.60  &    2.392   &   17.72   &    34.43  &     49.46   &    32.61   &    10.89  &     0  &    8.08 &  2.23e+12		\\
      93  &    1.73  &    3.131   &   8.783   &    10.84  &     18.91   &    13.82   &    2.563  &     0  &    12.8 &  1.76e+12	\\
      94  &    1.75  &    1.559   &   20.80   &    20.48  &     16.66   &    8.258   &    4.248  &     0  &    6.89 &  2.05e+12		\\
 \hline
\end{tabular}
\end{table*}

\bsp

\bibliographystyle{mn2e} 

\bibliography{mn2e_riguccini.bbl}

\label{lastpage}

\end{document}